\documentclass[12pt,fleqn]{article}
\usepackage[dvipdfmx]{graphicx}

\usepackage{amsmath,amssymb,amscd}
\usepackage{latexsym}
\usepackage{bm}
\usepackage{booktabs}
\usepackage{mathrsfs}
\usepackage{float}
\usepackage{fancybox}
\usepackage{braket}
\usepackage{ascmac}
\usepackage{fancybox,pifont,layout}
\usepackage{amsmath,amsthm}
\usepackage[cmintegrals]{newtxmath}
\usepackage{newtxtext}
\usepackage{mathrsfs}
\usepackage{mediabb}
\usepackage{cite}
\usepackage{url}
\usepackage{color}
\allowdisplaybreaks[3] 
\usepackage[square,sort&compress,comma,numbers]{natbib}
\renewcommand{\thefootnote}{(\roman{footnote})}

\makeatletter

\@addtoreset{equation}{section}
\makeatother

\begin{document}
\baselineskip=18pt
\begin{titlepage}
\begin{flushright}
{KOBE-TH-17-06}\\
{KIAS-P17067}
\end{flushright}
\vspace{1cm}
\begin{center}{\Large\bf 
Dynamical generation of fermion mass hierarchy\\
in an extra dimension
}
\end{center}
\vspace{0.5cm}
\begin{center}
Yukihiro Fujimoto$^{(a)}$\footnote{E-mail: \url{y-fujimoto@oita-ct.ac.jp}},
Takashi Miura$^{(b)}$\footnote{E-mail: \url{miura_takashi@jp.fujitsu.com}}\footnote{{This work is an achievement when the author belonged to Kobe university.}},
Kenji Nishiwaki$^{(c)}$\footnote{E-mail: \url{nishiken@kias.re.kr}},
Makoto Sakamoto$^{(d)}$\footnote{E-mail: \url{dragon@kobe-u.ac.jp}}
\end{center}
\begin{center}
${}^{(a)}${\it National Institute of Technology, Oita college, \\
Maki1666, Oaza, Oita 870-0152, Japan}\\
${}^{(b)}${\it Artificial Intelligence Laboratory, FUJITSU LABORATORIES LTD. 1-1,  \\
Kamikodanaka 4-chome, Nakahara-ku Kawasaki, 211-8588 Japan}\\
{${}^{(c)}${\it School of Physics, Korea Institute for Advanced Study,\\
Seoul 02455, Republic of Korea}}\\
${}^{(d)}${\it Department of Physics, Kobe University, \\
Kobe 657-8501, Japan}
\end{center}

\vspace{1cm}
\begin{abstract}
We propose a new mechanism to produce a fermion mass hierarchy dynamically {in a model with a single generation of fermions}. 
A five dimensional gauge theory on an interval with point interactions (zero-width branes) takes responsibility for realizing three generations and each massless zero mode localizes at boundaries of the segments on {the extra dimension.}
An extra-dimension coordinate-dependent vacuum expectation value of {a scalar field} makes large differences {in} overlap integrals of the localized zero modes and {then} {an exponential} fermion mass hierarchy can appear.
The positions of the point {interactions} control the magnitude of the fermion mass hierarchy and are determined by the minimization condition of the Casimir energy. 
As a result of the minimization of the Casimir energy, an exponential mass hierarchy appears dynamically. 
{We also discuss the stability of the extra dimension.}
\end{abstract}
\end{titlepage}
\newpage

\setcounter{footnote}{0}
\renewcommand{\thefootnote}{\arabic{footnote}\,}
\section{Introduction}
The standard model (SM), which provides an articulate description of the nature around TeV energy scale, was completed by the discovery of the Higgs boson \cite{:2012gk,:2012gu}. {However,} the SM still contains several mysteries and problems, which cannot be solved {within the context of the SM}. One is so-called the generation problem. The SM contains three {sets} of {quarks and leptons}, which {have {the} exact same quantum numbers} except for their Yukawa couplings. {Three generations} {were} introduced to the Kobayashi-Maskawa theory \cite{Kobayashi:1973fv} by hand {though the origin of the generations is not unveiled}.
Another {problem is on the fermion mass hierarchy}. Each generation of {the quarks and the charged leptons has {exactly the same quantum numbers} though their masses have an exponential hierarchy around $10^{5}$}. In the SM, the masses are generated by the Higgs mechanism and are determined by the dimensionless Yukawa couplings{; however,} there is no explanation to the question {of why so large a hierarchy }appears in the dimensionless parameters.

{Because of the above circumstance, various theories beyond the SM have been explored}. One possibility in the context of {four-dimensional (4d) gauge theory} is a {scenario} with non-compact gauge symmetry, {which can naturally produce the fermion mass hierarchy and three generations} \cite{Inoue:1994qz,Inoue:2000ia,Inoue:2003qi,Inoue:2007uj,Yamatsu:2012rj}. Another way is achieved by using extra dimensions. Extra dimension models with magnetic flux~\cite{Cremades:2004wa} {can lead to both of} the fermion mass hierarchy and three generations. {Magnetized orbifold models~\cite{Abe:2008fi,Abe:2008sx,Abe:2014noa,Abe:2014vza,Abe:2015yva,Fujimoto:2016zjs,Sakamura:2016kqv,Kobayashi:2016qag,Ishida:2017avx,Buchmuller:2017vho} {are}} also fascinating to discuss {within} the fermion flavor structure {and several} achievements {have} been investigated\footnote{{We can find an another geometric way to produce the generations in Refs.~\cite{Libanov:2000uf,Frere:2001ug,Frere:2003yv,Frere:2003ye,Frere:2010ah} in which {a topological structure of a vortex on a sphere plays} an important role.}}. However, some parameters of the models have to be chosen suitably by hand to make a fermion mass hierarchy. Moreover, in the case of extra dimension models, arguments for the stability of the extra dimension have mostly been postponed. Therefore, it is worth searching {the} dynamical generation mechanism of the fermion mass hierarchy and discussing the stability of the extra dimension, simultaneously.

In this paper, we propose a dynamical generation mechanism for the fermion mass hierarchy {in a model with a single generation of fermions {in five dimensions (5d)}. An interval extra dimension with point interactions \cite{Fujimoto:2011kf,Fujimoto:2012wv,Fujimoto:2013ki,Fujimoto:2014fka} {takes the responsibility} to produce the generations.\footnote{
{In 5d, to the best of our knowledge, 
the first proposal was given by the first manuscript of the series of our works~\cite{Fujimoto:2011kf}
with a concrete example where multiple chiral zero modes are generated 
from one five-dimensional fermion.}
}
Point interactions also play an important role to discuss the fermion mass hierarchy. In the previous model \cite{Fujimoto:2011kf,Fujimoto:2012wv,Fujimoto:2013ki,Fujimoto:2014fka}, the positions of  the point interactions, which affect to the fermion mass hierarchy, have been controlled {by hand}. On the other hand, in this paper, the positions of the point interactions are determined dynamically through the minimization of the Casimir energy~\cite{Ponton:2001hq,deAlbuquerque:2003qbk} (or say, Radion effective potential \cite{Garriga:2000jb,Goldberger:2000dv,Hofmann:2000cj,Abe:2014eia}). As a result, a large mass hierarchy appears dynamically in our model\footnote{See~\cite{Panico:2016ull} (also~\cite{Agashe:2016rle,DaRold:2017xdm}) for generating Yukawa hierarchies through multiple dynamical scales.}. {We also discuss the stability of the extra dimension from a Casimir energy point of view.}}

This paper is organized as follows. In {section}~\ref{sec:4d spectum of a 5d $U(1)$ gauge theory on an interval}, we review the 4d spectrum of a 5d $U(1)$ gauge theory on an interval extra dimension. {A general class of {boundary conditions (BCs)}, which is important to determine the 4d spectrum of the fields and phase structure of the symmetries, is derived for gauge, fermion and scalar {fields}. Using the knowledge of the general boundary conditions, we display the 4d spectrum at low energies and the profiles of the {mode functions} with respect to the extra dimension.} In {section}~\ref{sec.2-3}, we discuss {the} stability of the extra dimension. {Evaluating the {contribution} of each field, we investigate the extra dimension {length} dependence of the total Casimir energy.} {In} section~\ref{sec:Theory with point interactions}, a theory with point interactions is reviewed and {the 4d mass spectrum at low energies and the profiles of the mode functions are shown. In section~\ref{Semi-realistic}, using all the results, we construct {an $SU(2)\times U(1)$ model}, which can {lead to} the fermion mass hierarchy dynamically {with a single generation of fermions}. The minimization {of} the Casimir energy determines the positions of the point interactions, which are important parameters to {produce} the fermion mass hierarchy, and {leads} to the stability of the extra dimension. After that, we find that a fermion mass hierarchy {is realized} dynamically.} Section~\ref{sec:Conclusion and Discussion} is devoted to conclusion and discussion.
{In Appendix~\ref{sec:appendix}, we provide a self-contained review on the formulation of wave functions of a 5d fermion under the presence of one point interaction in the bulk part of an interval.}

\section{4d spectum of a 5d $U(1)$ gauge theory on an interval}\label{sec:4d spectum of a 5d $U(1)$ gauge theory on an interval}
In this section, we first summarize the results of allowed {boundary conditions, which {are} consistent with the {requirements from} the action principle, the gauge {invariance} and 4d Lorentz invariance, for gauge, fermion and scalar fields on an interval. The boundary conditions are crucially important to determine the 4d mass spectrum at low energies and also the phase structure of symmetries \cite{Sakamoto:1999yk,Sakamoto:1999ym,Sakamoto:1999iv,Ohnishi:2000hs,Hatanaka:2000zq,Sakamoto:2009hb,Hatanaka:2001kq,Hatanaka:2003mn,Lim:2005rc,Lim:2007fy,Lim:2008hi,Fujimoto:2011kf,Fujimoto:2012wv,Fujimoto:2013ki,Fujimoto:2014fka}. We then derive the 4d mass spectrum of the gauge and fermion fields, which are necessary to evaluate Casimir energies. We further show that the scalar field can possess a coordinate-dependent {vacuum expectation value (VEV)} on the extra dimension \cite{Sakamoto:1999yk,Sakamoto:1999ym,Sakamoto:1999iv,Fujimoto:2011kf,Fujimoto:2012wv,Fujimoto:2013ki,Fujimoto:2014fka}, which is found to be a crucial ingredient of our dynamical generation mechanism {for generating} a fermion mass hierarchy}.

\subsection{Consistent BCs for the fields}
In this subsection, we investigate the general class of BCs for an abelian gauge field, a fermion {field} and a scalar field on an interval, respectively. 
\subsubsection{BCs for {Abelian gauge, ghost{,} and anti-ghost fields}}
First, we start from the gauge field:
	\begin{align}
	S_{G}=\int d^4 x \int^{L}_{0}dy \Bigl[ -\frac{1}{4}F^{MN}F_{MN}-\frac{1}{2}(\partial^{\mu}A_{\mu}+\partial_{y}A_{y})^2 -i\bar{c}(\partial^{\mu}\partial_{\mu}+\partial_{y}^2)c\Bigr], \label{action_gauge}
	\end{align}
where
	\begin{align}
	F_{MN}=\partial_{M}A_{N}(x^{\mu},y)-\partial_{N}A_{M}(x^{\mu},y),\hspace{3em}(M,N=0,1,2,3,y).
	\end{align}
$x^{\mu}$ ($\mu=0,1,2,3$) denotes the {four-dimensional Minkowski-spacetime} coordinate and $y$ is the coordinate of the extra dimension with $0\leq y\leq L$. {Our choice of the 5d metric} is {$\eta_{MN}={\rm diag}(-1,1,1,1,1)$}. We {introduced} the second term as a gauge fixing term and the third term as a kinetic term of ghost fields. The general class of boundary conditions for the gauge field {is} obtained from the action principle:
	\begin{align}
	\delta S_{G}=0.
	\end{align}
We obtain the bulk field equation for $A_{M}$, together with the following surface term from the first term of the action after taking the variation.
	\begin{align}
	(\partial^{\mu}A_{y}-\partial_{y}A^{\mu})\delta A_{\mu}=0, \hspace{3em}{\rm at}\ y=0,L.
	\end{align}
Since the boundary condition $A_{\mu}=0$ at $y=0,L$ breaks 4d gauge symmetry explicitly, the general class of boundary conditions {consistent with the 4d gauge invariance} {is} given by the following{:}
	\begin{align}
	\left\{
	\begin{array}{l}
	\partial_{y}A_{\mu}=0,\\
	A_{y}=0,
	\end{array}
	\right. \hspace{3em}{\rm at}\ y=0,L. \label{Gauge_BCs}
	\end{align}
The BRST transformation leads us to BCs for the ghost field. The abelian gauge field $A_{M}$ and the ghost field $c$ have a relation with each other through the {Grassmann-odd} BRST transformation ${\bm \delta}_{B}$:
	\begin{align}
	{\bm \delta}_{B}A_{M}=\partial_{M}c.
	\end{align}
This fact implies that $\partial_{y} c$ ($c$) should obey the same boundary conditions as $A_{y}$ ($A_{\mu}$). Thus we obtain the BCs for the ghost as
	\begin{align}
	\partial_{y}c =0 \hspace{3em}{\rm at}\ y=0,L. \label{ghostBC}
	\end{align}
The boundary condition for the anti-ghost field $\bar{c}$ can be {derived} from the action principle for the third term of the action. The variation for the third term produces the following surface term{:}
	\begin{align}
	\bar{c}\partial_{y}(\delta c)-(\partial_{y}\bar{c})\delta c=0  \hspace{3em}{\rm at}\ y=0,L.
	\end{align}
Since $c(x,y)$ obeys the boundary conditions (\ref{ghostBC}), the following boundary condition should be imposed for the anti-ghost field $\bar{c}$:
	\begin{align}
	\partial_{y}\bar{c}=0 \hspace{3em}{\rm at}\ y=0,L. \label{anti-ghost_BC}
	\end{align}

\subsubsection{BCs for {fermion}}
Next, we consider the BCs for the fermion with adding the following action to eq.~(\ref{action_gauge}){:}
	\begin{align}
	S_{F}=\int d^4 x \int^{L}_{0}dy\ \overline{\Psi}(i\Gamma^{M}D_{M}+M_{F})\Psi, \label{Fermion_action}
	\end{align}
where
	\begin{align}
	D_{M}\Psi =(\partial_{M}-ieA_{M})\Psi,
	\end{align}
and $\Psi$ is a 5d 4-component Dirac spinor. $M_{F}$ is a bulk mass of the fermion and we take the gamma {matrix} $\Gamma^{M}$ as
	\begin{align}
	\Gamma^{\mu}&=\gamma^{\mu},\\
	\Gamma^{y}&=-i\gamma_{5}=\gamma^{0}\gamma^{1}\gamma^{2}\gamma^{3}.
	\end{align}
From the action principle $\delta S_F =0$, we obtain the following condition for the surface term:
	\begin{align}
	\overline{\Psi}\ \gamma_{5}\delta \Psi=0, \hspace{3em}{\rm at}\ y=0,L,
	\end{align}
with the 5d Dirac equation,
	\begin{align}
	i \gamma^{\mu}D_{\mu}\Psi +(\gamma_{5}D_{y}+M_{F})\Psi=0.
	\end{align}
In terms of the chiral spinors $\Psi_{R/L}$ ($\Psi =\Psi_{R}+\Psi_{L}$), which are defined as $\gamma_{5}\Psi_{R/L}=\pm \Psi_{R/L}$, we can rewrite the above equations as
	\begin{align}
	\overline{\Psi}_{L}\delta \Psi_{R}-\overline{\Psi}_{R}\delta\Psi_{L}=0, \label{fermion_surface}\hspace{3em}{\rm at}\ y=0,L,
	\end{align}
	\begin{align}
	i\gamma^{\mu}D_{\mu}\Psi_{R}+(-D_{y}+M_{F})\Psi_{L}=0,\\
	i\gamma^{\mu}D_{\mu}\Psi_{L}+(D_{y}+M_{F})\Psi_{R}=0.
	\end{align}
Since {boundary conditions} {which {consist} of a linear combination of $\Psi_{R}$ and $\Psi_{L}$} {break} the 4d Lorentz invariance, the condition{~(\ref{fermion_surface})} {should} be reduced to the form
	\begin{align}
	\overline{\Psi}_{L}\delta\Psi_{R}=0=\overline{\Psi}_{R}\delta \Psi_{L},
	\end{align}
which {leads to} the BCs:
	\begin{align}
	\Psi_{R}=0\hspace{1em}\text{or}\hspace{1em}\Psi_{L}=0.\hspace{3em}\text{at}\ y=0,L.
	\end{align}
We should note that under the BC $\Psi_{R}=0$ ($\Psi_{L}=0$) at boundaries, the 5d Dirac equation automatically determines the BC for $\Psi_{L}$ ($\Psi_{R}$) as
	\begin{align}
	\Psi_{R}=0 &\rightarrow (-D_{y}+M_{F})\Psi_{L}=0,\label{BCRtoL}\\
	\Psi_{L}=0  &\rightarrow (D_{y}+M_{F})\Psi_{R}=0. \label{BCLtoR}
	\end{align}
Thus we have the following four choices for the fermion BCs {\cite{Fujimoto:2011kf,Fujimoto:2012wv,Fujimoto:2013ki,Fujimoto:2014fka}}:
	\begin{align}
	\begin{array}{ll}
	\text{type-(I)}:\Psi_{R}(0)=0=\Psi_{R}(L),\\[0.2cm]
	\text{type-(II)}:\Psi_{L}(0)=0=\Psi_{L}(L),\\[0.2cm]
	\text{type-(III)}:\Psi_{R}(0)=0=\Psi_{L}(L),\\[0.2cm]
	\text{type-(IV)}:\Psi_{L}(0)=0=\Psi_{R}(L).
	\end{array}\label{sec2:Fermion_BCs}
	\end{align}

\subsubsection{BCs for {scalar} field}
Finally, we consider the general class of boundary conditions for a scalar field:
	\begin{align}
	S_{\Phi}=\int d^4 x \int^{L}_{0}dy \left[ \Phi^{\ast}(D^{M}D_{M} -M^2)\Phi -\frac{\lambda}{2}(\Phi^{\ast}\Phi)^2\right],\label{Scalar_action}
	\end{align}
where
	\begin{align}
	D_{M}\Phi=(\partial_{M}-ie'A_{M})\Phi,
	\end{align}
and $\Phi(x,y)$ denotes a 5d complex scalar {field. As} {in} the previous cases, we obtain the surface term from the action principle $\delta S_{\Phi}=0$:
	\begin{align}
	\Phi^{\ast}D_{y}\delta \Phi -(D_{y}\Phi)^{\ast}\delta\Phi=0, \hspace{3em}{\rm at}\ y=0,L. \label{surface_term}
	\end{align}
Under the infinitesimal special variation $\delta \Phi =\varepsilon \Phi$, we can rewrite the above surface term as 
	\begin{align}
	|\Phi -iL_{0}D_{y}\Phi|^{2}=|\Phi+iL_{0}D_{y}\Phi|^{2}\hspace{3em}{\rm at}\ y=0,L,
	\end{align}
where $L_{0}$ is an arbitral non-zero real constant, which possesses mass dimension {$-1$}. The above equation implies that $\Phi -iL_{0}D_{y}\Phi$ and $\Phi+iL_{0}D_{y}\Phi$ have a difference only up to a phase {at the boundaries}:
	\begin{align}
	&\Phi -iL_{0}(D_{y}\Phi)=e^{i\theta_{0}}(\Phi+iL_{0}D_{y}\Phi)\hspace{3em}{\rm at }\ y=0,\\
	&\Phi -iL_{0}(D_{y}\Phi)=e^{i\theta_{L}}(\Phi+iL_{0}D_{y}\Phi)\hspace{3em}{\rm at }\ y=L.
	\end{align}
{With} $L_{+}\equiv L_{0}\cot \frac{\theta_{0}}{2}$ and $L_{-}\equiv -L_{0}\cot\frac{\theta_{L}}{2}$, we obtain the general class of BCs for the scalar field {\cite{Fujimoto:2011kf,Fujimoto:2012wv,Fujimoto:2013ki,Fujimoto:2014fka}},
	\begin{align}
	\left\{
	\begin{array}{l}
	\Phi(0)+L_{+}D_{y}\Phi(0)=0,\\
	\Phi(L)-L_{-}D_{y}\Phi(L)=0,
	\end{array}
	\right.\hspace{3em}(-\infty \leq L_{\pm}\leq +\infty).\label{Robin_BCs}
	\end{align}
{These} boundary conditions are known as the Robin boundary condition. Note that the derived Robin boundary condition satisfies the condition~(\ref{surface_term}) under the assumption that $\Phi$ and $\delta \Phi$ satisfy the same boundary condition. 

We should emphasize that all derived boundary conditions (\ref{Gauge_BCs}), (\ref{ghostBC}), (\ref{anti-ghost_BC}), (\ref{sec2:Fermion_BCs}), (\ref{Robin_BCs}) are consistent with the 5d gauge invariance.

\subsection{4d spectrum}
In the previous subsection, we {investigated} the general class of BCs for each field. Now, we derive the 4d spectrum of the gauge field and the fermion field under the derived boundary conditions, respectively. For the scalar field, we only investigate the vacuum expectation value for {our} purpose.

\subsubsection{4d spectrum of Abelian {gauge, ghost{,} and anti-ghost fields}}
\begin{figure}[h]
	\begin{center}
	\includegraphics[width=6cm]{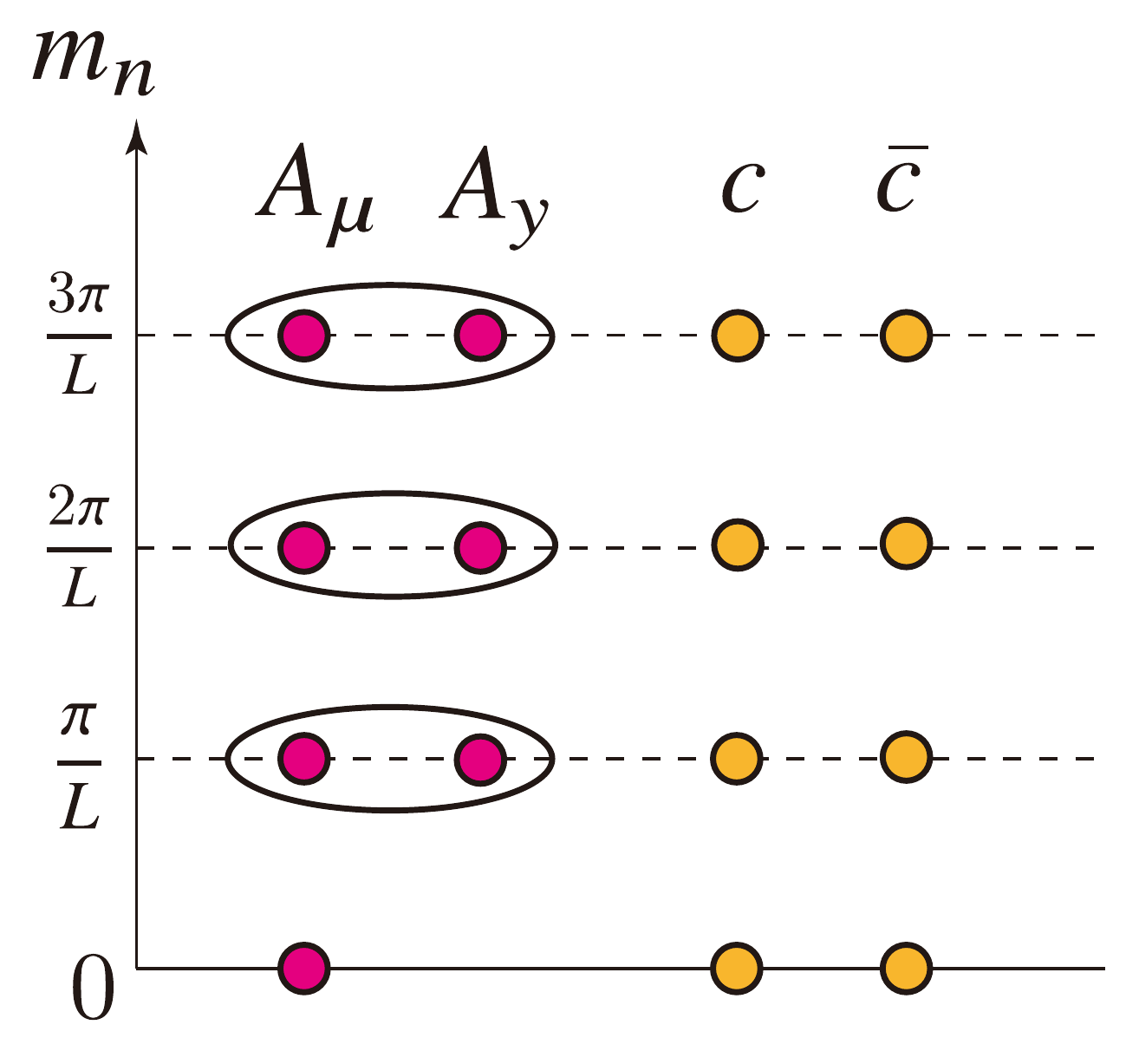}
	\caption{A schematic figure of the 4d spectrum of the abelian gauge field with ghosts on an interval. Each black oval pair indicates a QM-SUSY pair to make a mass term}
	\label{fig.4d-gauge}
	\end{center}
	\end{figure}

First, we start from the abelian {gauge, the ghost{,} and the anti-ghost fields}. The action and the boundary conditions are given by eq.~(\ref{action_gauge}) and eqs.~(\ref{Gauge_BCs}),~(\ref{ghostBC}),~(\ref{anti-ghost_BC}).
The action $S_{G}$ can be rewritten as 	
	\begin{align}
	S_{G}=\int d^4 x\int^{L}_{0}dy\ \left[ \frac{1}{2}A^{\mu}(\partial^{\nu}\partial_{\nu}+\partial_{y}^2)A_{\mu}+\frac{1}{2}A_{y}(\partial^{\mu}\partial_{\mu}+\partial_{y}^2)A_{y}-i\bar{c}(\partial^{\mu}\partial_{\mu}+\partial_{y}^2)c\right].\label{action_gauge2}
	\end{align}
To obtain the 4d spectrum, we expand the fields as follows:
	\begin{align}
	A_{\mu}(x,y)=\sum_{n}A^{(n)}_{\mu}(x)f_{n}(y),\\
	A_{y}(x,y)=\sum_{n}A^{(n)}_{y}(x)g_{n}(y),\\
	c(x,y)=\sum_{n}c^{(n)}(x)\Xi_{n}(y),\\
	\bar{c}(x,y)=\sum_{n}\bar{c}^{(n)}(x)\Xi_{n}(y),
	\end{align}
where $\{f_{n}(y)\}$ $\Bigl( \{g_{n}(y)\}\Bigr)$ are eigenfunctions of the Hermitian operator ${\cal D}^{\dagger}{\cal D}$ (${\cal D}{\cal D}^{\dagger}$):
	\begin{align}
	\left\{
	\begin{array}{l}
	{\cal D}^{\dagger}{\cal D}f_{n}(y)=m_{n}^2 f_{n}(y),\\[0,3cm]
	{\cal D}{\cal D}^{\dagger}g_{n}(y)=m_{n}^2 g_{n}(y),
	\end{array}
	\right.
	\end{align}
and we defined ${\cal D}$ and ${\cal D}^{\dagger}$ as
	\begin{align}
	{\cal D}&\equiv \partial_{y},\\
	{\cal D}^{\dagger} &\equiv -\partial_{y}.
	\end{align}
$\{\Xi_{n}(y)\}$ are eigenfunctions of the Hermitian operator $(-\partial_{y}^2)$,
	\begin{align}
	-\partial_{y}^2 \Xi_{n}(y)=m_{n}^2 \Xi_{n}(y).
	\end{align}
Note that $\{ f_{n}\}$, $\{g_{n}\}${,} and $\{\Xi_{n}\}$ form {complete sets}, respectively, and can obey the {orthonormal} relations:
	\begin{align}
	\int^{L}_{0}dy\ f_{n}^{\ast}(y)f_{m}(y)=\delta_{n,m},\\
	\int^{L}_{0}dy\ g_{n}^{\ast}(y)g_{m}(y)=\delta_{n,m},\\
	\int^{L}_{0}dy\ \Xi^{\ast}_{n}(y)\Xi_{m}(y)=\delta_{n,m}.
	\end{align}
Furthermore, $\{f_{n}\}$ and $\{g_{n}\}$ satisfy the {quantum-mechanical} supersymmetry (QM-SUSY) relations \cite{Lim:2005rc,Lim:2007fy,Lim:2008hi,Ohya:2010wf,Nagasawa:2011mu,Sakamoto:2012ew}, 
	\begin{align}
	\left\{
	\begin{array}{l}
	{\cal D}f_{n}(y)=m_{n}g_{n}(y),\\
	{\cal D}^{\dagger}g_{n}(y)=m_{n}f_{n}(y).
	\end{array}
	\right.
	\end{align}
Under the BCs~(\ref{Gauge_BCs}),~(\ref{ghostBC}),~(\ref{anti-ghost_BC}), we can derive the explicit {forms} of $\{f_{n}\}$, $\{g_{n}\}$ and $\{\Xi_{n} \}$ with the mass eigenvalue $m_{n}$ as 
	\begin{align}
	\begin{array}{ll}
	f_{0}=\sqrt{\frac{1}{L}},&\\
	f_{n}=\sqrt{\frac{2}{L}}\cos\left(\frac{n\pi}{L}y\right),&\hspace{3em}(n=1,2,3,\cdots),\\
	g_{n}=-\sqrt{\frac{2}{L}}\sin\left(\frac{n\pi}{L}y\right),&\hspace{3em}(n=1,2,3,\cdots),\\
	\Xi_{0}=\sqrt{\frac{1}{L}},&\\
	\Xi_{n}=\sqrt{\frac{2}{L}}\cos\left(\frac{n\pi}{L}y\right),&\hspace{3em}(n=1,2,3,\cdots),\\
	{{m_{n}}=\frac{n\pi}{L}},&\hspace{3em}(n=0,1,2,\cdots).
	\end{array}
	\end{align}
Substituting the {above expansions} into the action~(\ref{action_gauge2}) and executing the integration with respect to the extra dimension, we obtain the following reduced action.
	\begin{align}
	S_{G}&=\int d^4 x \biggl[ \frac{1}{2}A^{(0)}_{\mu}\eta^{\mu\nu}(\partial^{\alpha}\partial_{\alpha})A^{(0)}_{\nu} +\sum^{\infty}_{n=1}\frac{1}{2}A^{(n)}_{\mu}\eta^{\mu\nu}(\partial^{\alpha}\partial_{\alpha}-m_{n}^2)A_{\nu}^{(n)}+\sum^{\infty}_{n=1}\frac{1}{2}A_{y}^{(n)}(\partial^{\alpha}\partial_{\alpha}-m_{n}^2)A_{y}^{(n)}\nonumber\\
	&\hspace{5em}-i\bar{c}^{(0)}(\partial^{\alpha}\partial_{\alpha})c^{(0)}-i\sum^{\infty}_{n=1}\bar{c}^{(n)}(\partial^{\alpha}\partial_{\alpha}-m_{n}^2)c^{(n)}\biggr].
	\end{align}
A schematic figure of the 4d spectrum is depicted in Figure~\ref{fig.4d-gauge}.

\subsubsection{4d spectrum of fermion}
\begin{center}
	\begin{figure}[h]
	\begin{center}
		\begin{minipage}{0.9\textwidth}
		\begin{center}
		\scalebox{0.3}{\includegraphics{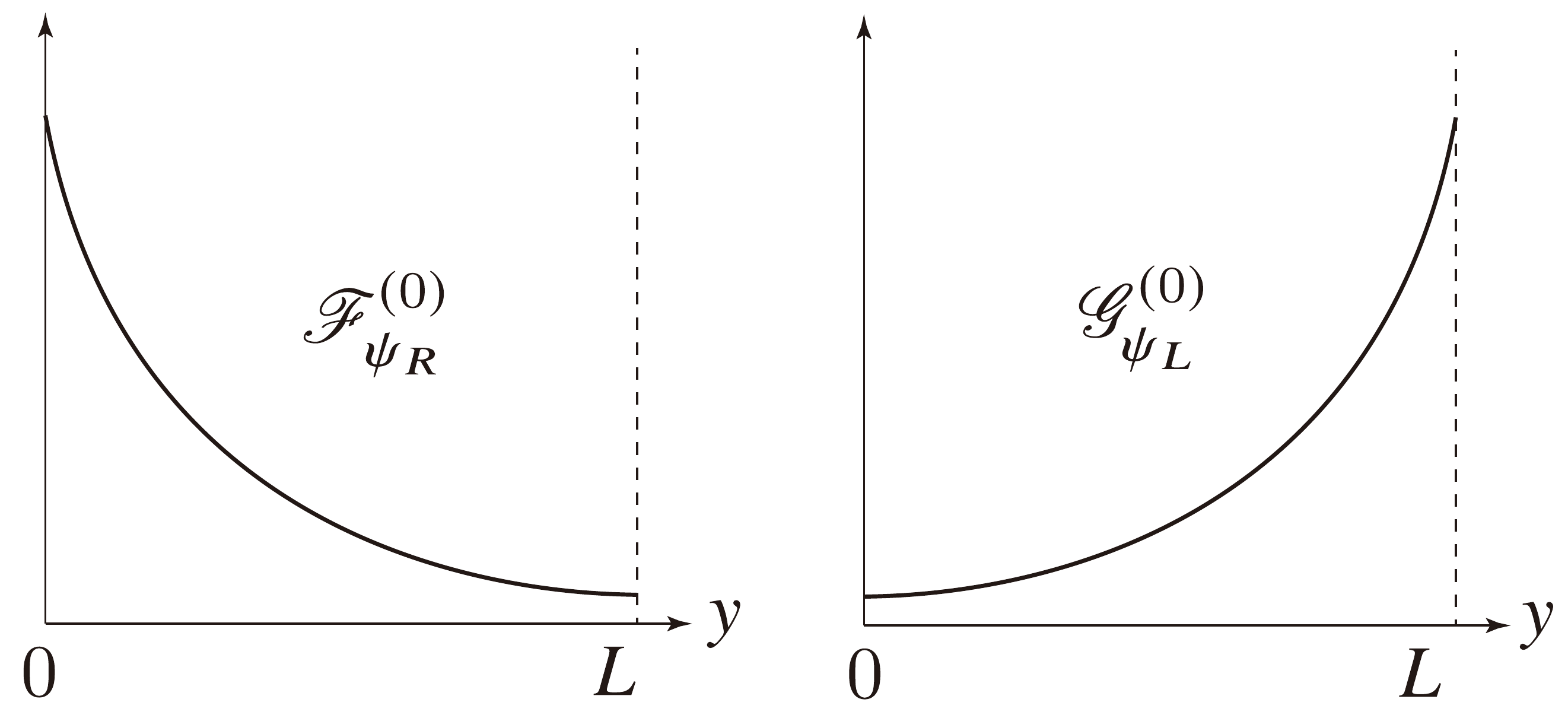}}
		\par
		{\footnotesize{(i) {Schematic figures} of $\mathscr{F}^{(0)}_{\psi_R}$ and $\mathscr{G}^{(0)}_{\psi L}$ in the case of $M_{F}>0$. $\mathscr{F}^{(0)}_{\psi_R}$ ($\mathscr{G}^{(0)}_{\psi_L}$) localizes to the boundary point $y=0$ ($y=L$).}}
		\end{center}
		\end{minipage}
	\end{center}
	\begin{center}
		\begin{minipage}{0.9\textwidth}
		\begin{center}
		\scalebox{0.3}{\includegraphics{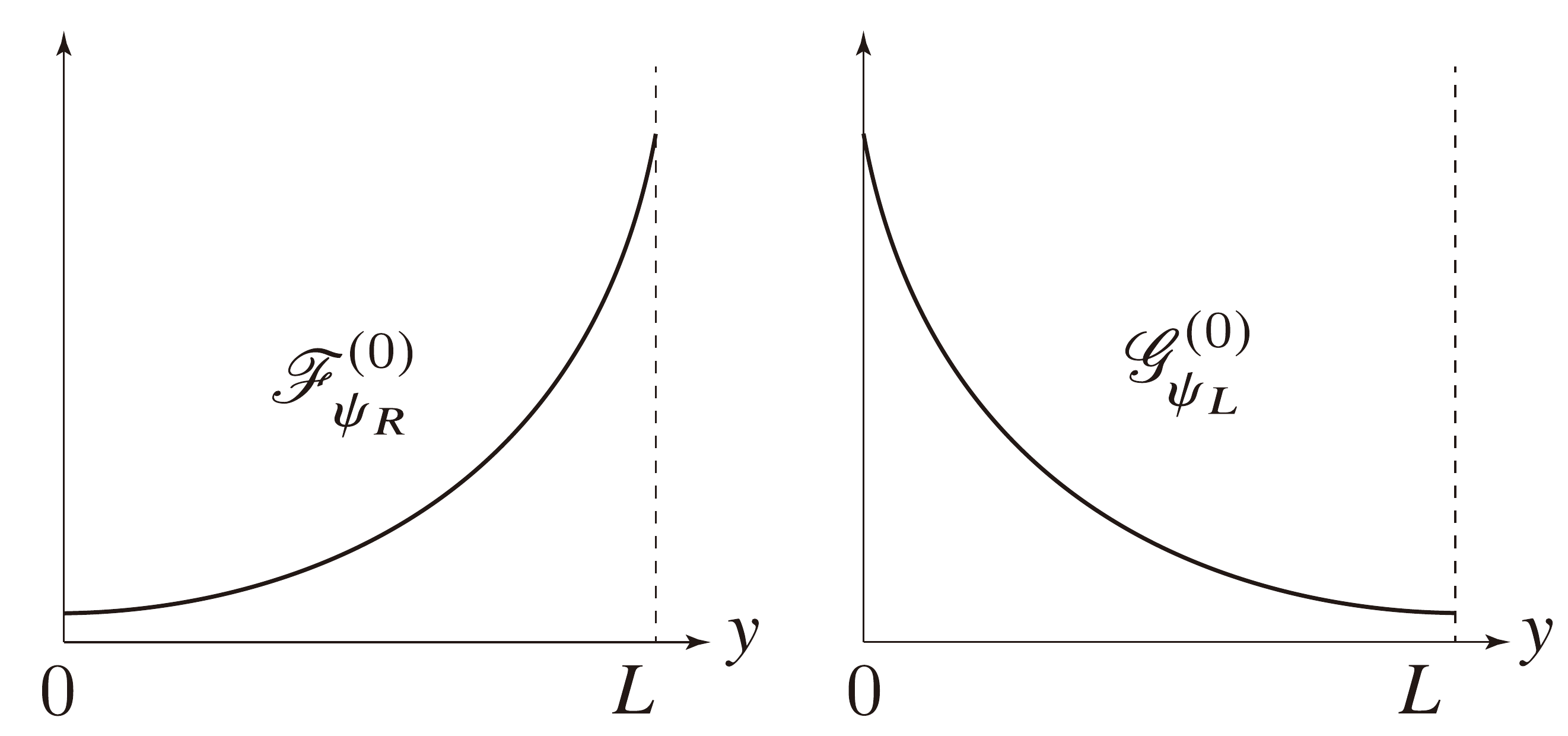}}
		\par
		\vspace{0.0cm}
		{\footnotesize{(ii) {Schematic figures} of $\mathscr{F}^{(0)}_{\psi_R}$ and $\mathscr{G}^{(0)}_{\psi_L}$ in the case of $M_{F}<0$. $\mathscr{F}^{(0)}_{\psi_R}$ ($\mathscr{G}^{(0)}_{\psi_L}$) localizes to the boundary point $y=L$ ($y=0$).}}
		\end{center}
		\end{minipage}
	\end{center}
	\caption{{Schematic figures} of chiral massless {fermion} zero mode solutions.}
	\label{fig.zero-modes}
	\end{figure}
	\end{center}

\begin{figure}[h]
		\begin{center}
		\begin{minipage}{0.3\textwidth}
		\begin{center}
		\scalebox{0.45}{\includegraphics{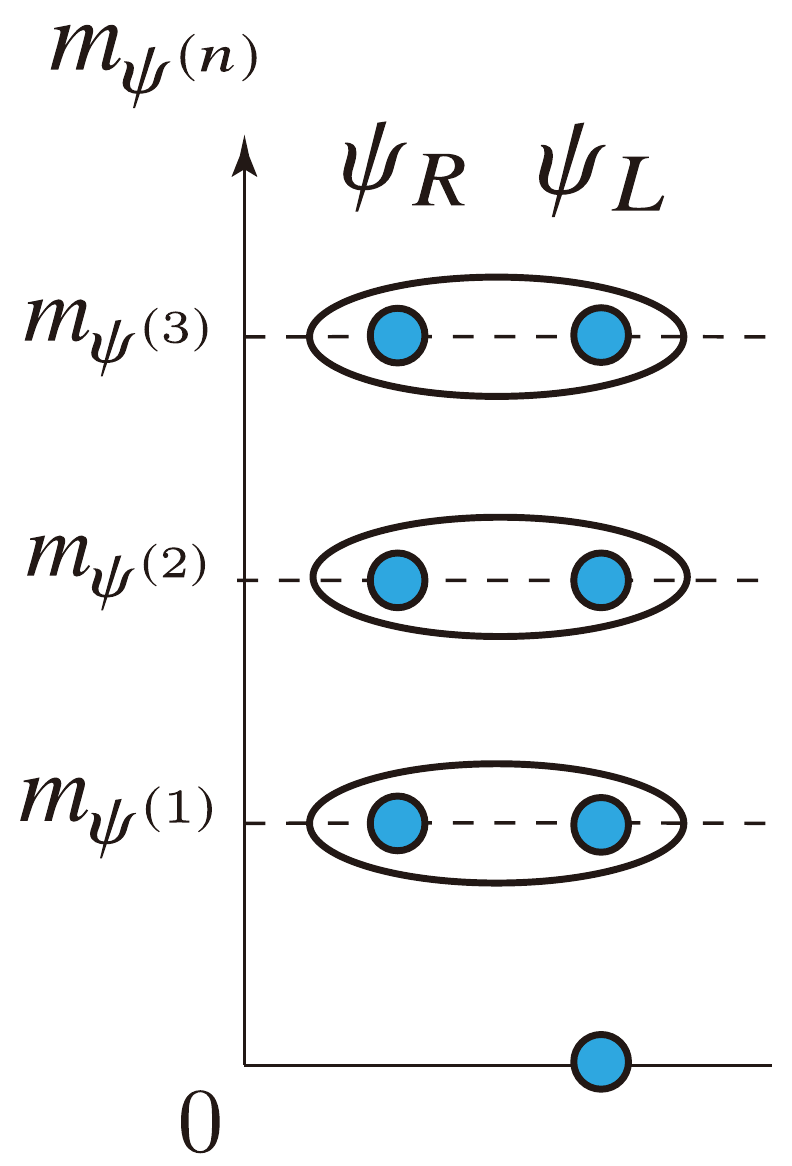}}
		\par
		\vspace{0.0cm}
		{\footnotesize{{Type-(I) case}}}
		\end{center}
		\end{minipage}
		\begin{minipage}{0.3\textwidth}
		\begin{center}
		\scalebox{0.45}{\includegraphics{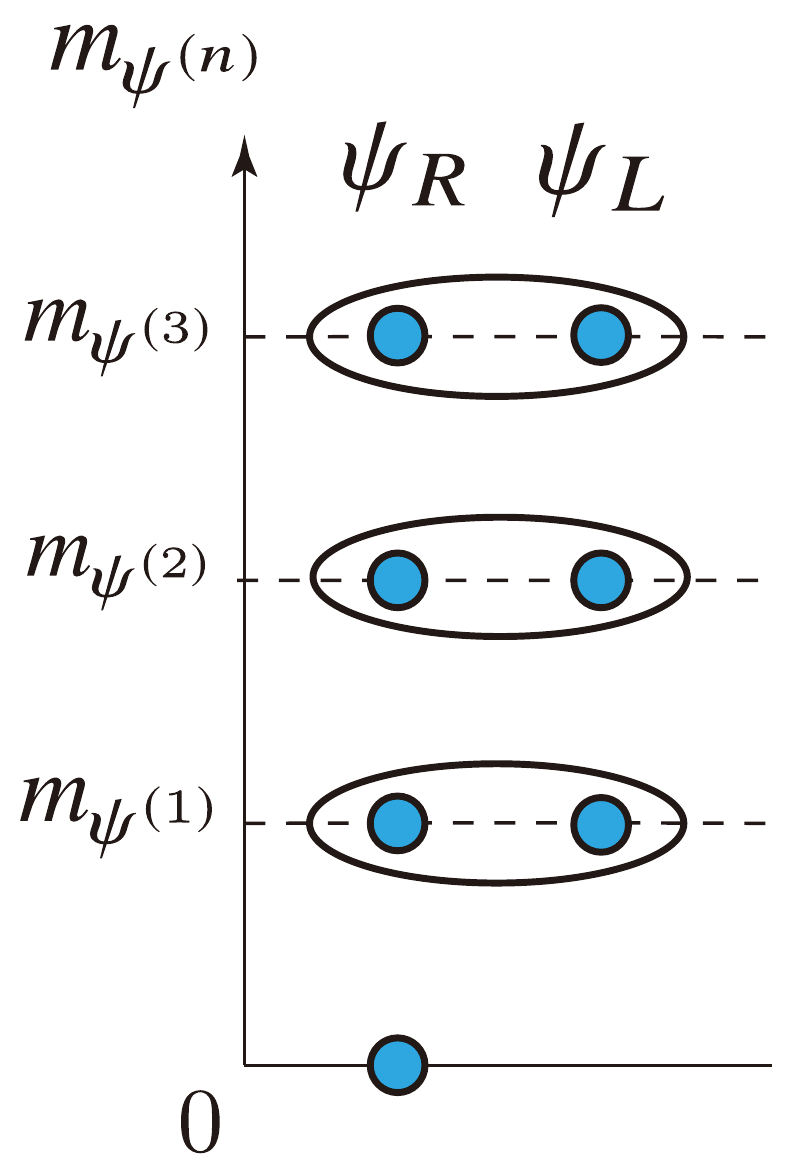}}
		\par
		\vspace{0.0cm}
		{\footnotesize{{Type-(II) case}}}
		\end{center}
		\end{minipage}
		\end{center}
		\caption{A typical mass spectrum of the fermion on an interval. Each black oval pair indicates a QM-SUSY pair to make a mass term.}
		 \label{fig.chap2-4d-fermion}
	\end{figure}

Second, we investigate the 4d spectrum of the fermion on an interval. The action and BCs are given by eq.~(\ref{Fermion_action}) and {eq.~(\ref{sec2:Fermion_BCs}).
To evaluate the 4d spectrum of the fermion, we expand the fermion as
	\begin{align}
	\Psi(x,y)&=\Psi_{R}(x,y)+\Psi_{L}(x,y)\nonumber\\
	&=\sum_{n}\psi^{(n)}_{R}(x)\mathscr{ F}^{(n)}_{\psi_R}(y)+\sum_{n}\psi^{(n)}_{L}(x)\mathscr{ G}^{(n)}_{\psi_L}(y),
	\end{align}
}
where $\{ \mathscr{F}^{(n)}_{\psi_R}\}$ ($\{\mathscr{G}^{(n)}_{\psi_L}\}$) {are eigenfunctions} of the {Hermitian} operator $\mathscr{D}^{\dagger}\mathscr{D}$ ($\mathscr{D}\mathscr{D}^{\dagger}$):
	\begin{align}
	\left\{
	\begin{array}{l}
	\mathscr{D}^{\dagger}\mathscr{D}\mathscr{F}^{(n)}_{\psi_R}(y)=m_{\psi^{(n)}}^2 \mathscr{F}^{(n)}_{\psi_R}(y),\\[0.2cm]
	\mathscr{D}\mathscr{D}^{\dagger}\mathscr{G}^{(n)}_{\psi_L}(y)=m_{\psi^{(n)}}^2 \mathscr{G}^{(n)}_{\psi_L}(y),
	\end{array}
	\right.\label{Fermion_eigenvalue-equation}
	\end{align}
and {form} {complete sets}. In the above, the operators $\mathscr{D}$ and $\mathscr{D}^{\dagger}$ are defined as
	\begin{align}
	\mathscr{D}&\equiv \partial_{y}+M_{F},\\
	\mathscr{D}^{\dagger}&\equiv -\partial_{y}+M_{F}.
	\end{align}
Furthermore, $\{ \mathscr{F}^{(n)}_{\psi_R}\}$ and $\{\mathscr{G}^{(n)}_{\psi_L} \}$ satisfy the QM-SUSY relations:
	\begin{align}
	\left\{
	\begin{array}{l}
	\mathscr{D}\mathscr{F}^{(n)}_{\psi_R}(y)=m_{\psi^{(n)}}\mathscr{G}^{(n)}_{\psi_L}(y),\\[0.2cm]
	\mathscr{D}^{\dagger}\mathscr{G}^{(n)}_{\psi_L}(y)=m_{\psi^{(n)}}\mathscr{F}^{(n)}_{\psi_R}(y).
	\end{array}
	\right.\label{eq:QM-SUSY relations}
	\end{align}
We can obtain the explicit {forms} of the wavefunctions after we solve the eigenvalue {equations} (\ref{Fermion_eigenvalue-equation}) {while taking into account} the BCs (\ref{sec2:Fermion_BCs}). However, we {here} concentrate on the existence of {a} chiral massless zero-mode and the form of its {wavefunction.} {Zero-mode solutions are} obtained from the QM-SUSY {relations (\ref{eq:QM-SUSY relations}) with} $m_{\psi^{(0)}}=0$:
	\begin{align}
	\mathscr{D}\mathscr{F}^{(0)}_{\psi_R}&=0, \label{ZEROsolutionF}\\
	\mathscr{D}^{\dagger}\mathscr{G}^{(0)}_{\psi_L}&=0.\label{ZEROsolutionG}
	\end{align}
The {solutions} of the above {equations would be given} as follows:
	\begin{align}
	\mathscr{F}^{(0)}_{\psi_R}(y)&=\sqrt{\frac{2M_{F}}{1-e^{-2M_{F}L}}}e^{-M_{F}y},\label{f0R}\\
	\mathscr{G}^{(0)}_{\psi_L}(y)&=\sqrt{\frac{2M_{F}}{e^{2M_{F}L}-1}}e^{M_{F}y}.\label{g0L}
	\end{align}
{Schematic figures} of the zero-mode {solutions are} depicted in Figure~\ref{fig.zero-modes}. The zero-mode solution $\mathscr{F}^{(0)}_{\psi_R}$ $\Bigl(\mathscr{G}^{(0)}_{\psi_L}\Bigr)$ {localizes} to the boundary $y=0$ ($y=L$) in the case of $M_{F}>0$ and localizes to $y=L$ ($y=0$) in the case of $M_{F}<0$.
It should be emphasized that the {zero-mode} solutions $(\ref{f0R})$ $\bigl((\ref{g0L})\bigr)$ {are} consistent only with the type-(II) $\bigl($type-(I)$\bigr)$ BC given in (\ref{sec2:Fermion_BCs}) because of (\ref{BCRtoL}) and (\ref{BCLtoR}){, respectively.} Therefore we will concentrate on {the} {type-(I) and type-(II)} BCs in the following. The mass spectrum of both type-(I) and type-(II) is {given by}
	\begin{align}
	m_{\psi^{(0)}}&=0,\\
	{m_{\psi^{(n)}}}&{= \sqrt{\left(\frac{n\pi}{L}\right)^2 +M_{F}^2}} , \hspace{4em}(n=1,2,3,\cdots ).
	\end{align}
Inserting the mode expansions into the action and using the {orthonormal} relations of the mode functions, we have
	\begin{align}
	S_{F}=\int d^4 x \left\{ {\cal L}_{m=0}+\sum^{\infty}_{n=1}\overline{\psi^{(n)}}(x)\Bigl(i\gamma^{\mu}\partial_{\mu}+m_{n}\Bigr)\psi^{(n)}(x)\right\},
	\end{align}
where
	\begin{align}
	{\cal L}_{m=0}&= \left\{ 
					\begin{array}{l}
					\overline{\psi^{(0)}_{L}}(x) (i\gamma^{\mu}\partial_{\mu})\psi^{(0)}_{L}(x), \hspace{4em}\text{{for}\quad type-(I)},\\[0.3cm]
					\overline{\psi^{(0)}_{R}}(x) (i\gamma^{\mu}\partial_{\mu})\psi^{(0)}_{R}(x), \hspace{4em}\text{{for}\quad type-(II)},
					\end{array}
					\right.
	\end{align}
and $\psi^{(n)}=\psi^{(n)}_R +\psi^{(n)}_L$.
A typical spectrum of the fermion is depicted {in Figure~\ref{fig.chap2-4d-fermion}}. A chiral massless zero mode {exists} in the case of both type-(I) and type-(II).

\subsubsection{Vacuum expectation value of the scalar}
Finally, we comment on the vacuum expectation value of the scalar field. The action and the BCs are given by eq.~(\ref{Scalar_action}) and eq.~(\ref{Robin_BCs}).
It was found in Refs.~\cite{Fujimoto:2011kf,Fujimoto:2012wv} that under the Robin boundary condition (\ref{Robin_BCs}), $\Phi(x,y)$ can {possess} a non-vanishing vacuum expectation value $\langle \Phi(x,y)\rangle=\phi(y)$ with the form
	\begin{align}
	\phi(y) =\dfrac{\frac{M}{\sqrt{\lambda}}(\sqrt{1+X}-1)^{\frac{1}{2}}}{{\rm cn}\left[M(1+X)^{\frac{1}{4}}(y-y_{0}), \sqrt{\frac{1}{2}(1+\frac{1}{\sqrt{1+X}}})\right]},
	\end{align}
with
	\begin{align}
	X=\frac{4\lambda |Q|}{M^4}.
	\end{align}
${\rm cn}(y,a)$ is the Jacobi's elliptic function, {and $y_0$, $Q$} are constants which are determined by the parameters $L_{\pm}$ of the Robin BCs. {Choosing} {suitable values} of $L_{\pm}$, {we can approximately {take} the form of the scalar VEV $\phi(y)$ as 
	\begin{align}
	\phi(y)\sim{\cal A}e^{My},\label{y-dependentVEV}
	\end{align}
}
where ${\cal A}$ is a {constant with} mass dimension $\frac{3}{2}$.

\section{Casimir energy and stability of the extra dimension}\label{sec.2-3}
In the previous section, we succeeded {in obtaining} the 4d spectrum of the fields with the specified BCs. {Taking the result into account,} we evaluate the Casimir energy $E[L]$ as a function of the {length} $L$ of the extra dimension {and show that the minimization of the Casimir energy provides a mechanism to stabilize the extra dimension}.

For our purpose, we only {concentrate} on the gauge {and fermion field contributions to the Casimir energy} {while} ignoring the effect of the scalar field in this paper. {We summarize the action and the BCs which we consider,}
	\begin{align}
	S&=S_G +S_F,\\
	S_{G}&=\int d^4 x \int^{L}_{0}dy \Bigl[ -\frac{1}{4}F^{MN}F_{MN}-\frac{1}{2}(\partial^{\mu}A_{\mu}+\partial_{y}A_{y})^2 -i\bar{c}(\partial^{\mu}\partial_{\mu}+\partial_{y}^2)c\Bigr], \\
	S_{F}&=\int d^4 x \int^{L}_{0}dy\ \overline{\Psi}(i\Gamma^{\mu}\partial_{\mu}+i\Gamma^{y}\partial_{y}+M_{F})\Psi,
	\end{align}
	\begin{align}
	\left\{
	\begin{array}{l}
	\partial_{y}A_{\mu}=0,\\
	A_{y}=0,
	\end{array}
	\right. \hspace{3em}{\rm at}\ y=0,L{,} \label{Sec2Gauge_BCs}
	\end{align}
	\begin{align}
	\partial_{y}c=0=\partial_{y}\bar{c} \hspace{3em}{\rm at}\ y=0,L, \label{Sec2ghost_BC}
	\end{align}
	\begin{align}
	\begin{array}{ll}
	\text{type-(I)}:\Psi_{R}(0)=0=\Psi_{R}(L),\\[0.2cm]
	\text{type-(II)}:\Psi_{L}(0)=0=\Psi_{L}(L){.}\label{Fermion_BCs}
	\end{array}
	\end{align}
	We {focus} on the situation in which a chiral massless zero mode exists. As an {example}, we consider {the type-(II) BC} {first}.
	To evaluate the Casimir energy, we examine the partition function $Z[L]$.
	\begin{align}
	Z[L]=\int [dA_{\mu}\, dA_y\, {d\Psi\, d\bar{\Psi}}\, dc\, d\bar{c} ]\,e^{iS}.
	\end{align}
	The gauge field part of the partition function {reads}
	\begin{align}
	Z_{G}[L]&=\int [dA_\mu dA_y dc d\bar{c}]\, e^{iS_{G}}\\
	&\propto{\rm exp}\left[ i\int d^4 x \Biggl\{ i\int \frac{d^4 p}{(2\pi)^4}\,\biggl({\rm ln}\, p^{\mu}p_{\mu}+\frac{3}{2}\sum^{\infty}_{n=1}{\rm ln}\, (p^{\mu}p_{\mu}+m_{n}^2)\biggr)\Biggr\}\right].
	\end{align}
After moving to the Euclidian space, we obtain the Casimir energy of the gauge field:
	\begin{align}
	Z^{{\rm Euclid}}_{G}[L]\propto{\rm exp}\left[ -E^{\rm U(1)}[L]\int d^4 x_{E}\right],
	\end{align}
where 
	\begin{align}
	E^{\rm U(1)}[L]&= \int \frac{d^4 p_{E}}{(2\pi)^4}\left[ {\rm ln}\, p_{E}^2 +\frac{3}{2}\sum^{\infty}_{n=1}{\rm ln}\, (p_{E}^2 +m_{n}^2)\right]\nonumber\\
	&= \int \frac{d^4 p_{E}}{(2\pi)^4}\left[ \frac{1}{4}\,{\rm ln}\, p_{E}^2 +\frac{3}{4}\sum^{\infty}_{n=-\infty}{\rm ln} \Bigl\{ p_{E}^2 +\left(\frac{n\pi}{L}\right)^2\Bigr\}\right].
	\end{align}
and $p_{E}^2 =(p^0_E)^2+(p^1_E)^2+(p^2_E)^2+(p^3_E)^2$.
{For further concrete discussions,} we {divide} $E^{\rm U(1)}[L]$ into two parts:
	\begin{align}
	E^{\rm U(1)}[L]&=E^{\rm U(1)}_{{\rm part1}}[L]+E^{\rm U(1)}_{{\rm part2}}[L],\\
	E^{\rm U(1)}_{{\rm part1}}[L]&= \int \frac{d^4 p_{E}}{(2\pi)^4}\ \frac{1}{4}\,{\rm ln}\, p_{E}^2,\\
	E^{\rm U(1)}_{{\rm part2}}[L]&= \int \frac{d^4 p_{E}}{(2\pi)^4}\ \frac{3}{4}\sum^{\infty}_{n=-\infty} {\rm ln}\, \biggl\{ p_{E}^2 +\left(\frac{n\pi}{L}\right)^2\biggr\}
	\end{align}
{Now we find that $E^{\rm U(1)}_{\rm part1}[L]$ has no $L$-dependence. {Our interest is} only in the $L$-dependence of the Casimir energy $E^{\rm U(1)}[L]$ so that we simply ignore this part. We emphasize that this part actually does not affect any results of the $L$-dependence of the Casimir energy $E^{\rm U(1)}[L]$.} On the other hand, $E^{\rm U(1)}_{\rm part2}[L]$ has {an} $L$-dependence and {plays} a crucial role when we discuss the $L$-dependence of the total Casimir energy. 
{By using the formulas}
	\begin{align}
	&-{\rm ln}\, A=\frac{d}{ds}A^{-s}\biggr|_{s=0},\label{formula1}\\
	&A^{-s}=\frac{1}{\Gamma (s)}\int^{\infty}_{0} dt\ t^{s-1}e^{-At},\label{formula2}\\
	&\frac{d}{ds}\frac{t^s}{\Gamma (s)}\Biggr|_{s=0}=1,\label{formula3}
	\end{align}
with the Gamma function $\Gamma (s)=\int^{\infty}_{0}dt\, t^{s-1}e^{-t}$, we can rewrite $E^{\rm U(1)}_{\rm part2}[L]$ as
	\begin{align}
	E^{\rm U(1)}_{\rm part2}[L]=-\frac{3}{4}\cdot\frac{1}{16\pi^2}\sum^{\infty}_{n=-\infty}\int^{\infty}_{0}dt\, t^{-3} e^{-(\frac{n\pi}{L})^2 t}.
	\end{align}
The Poisson summation {formula}
	\begin{align}
	\sum^{\infty}_{n=-\infty}e^{-(\frac{n\pi}{L})^2 t}=\sum^{\infty}_{w=-\infty}\frac{L}{\sqrt{\pi  t}}e^{-\frac{w^2 L^2}{t}},\label{poisson}
	\end{align}
will help us to move on. Here, the index $w$ is {an} integer which {represents} the winding number. {By} {utilizing} the Poisson summation formula, we obtain
	\begin{align}
	E^{\rm U(1)}_{\rm part2}[L]=-\frac{3L}{64\pi^{5/2}}\sum^{\infty}_{w=-\infty}\int^{\infty}_{0}dt\, {t}^{-\frac{7}{2}}e^{-\frac{w^2 L^2}{t}},
	\end{align}
and find that $E^{\rm U(1)}_{\rm part2}[L]$ contains a UV-divergence {when} $t\rightarrow 0$. To remove this UV-divergence, we define the regularized total Casimir energy as
	\begin{align}
	\frac{1}{L}E^{\rm U(1)}_{\rm part2}[L]_{\rm reg.}\equiv \frac{1}{L}E^{\rm U(1)}_{\rm part2}[L] -\frac{1}{L}E^{\rm U(1)}_{\rm part2}[L]\Biggr|_{{L\to\infty}} .
	\end{align}
{{We note that this} regularization {is equivalent simply to removing} the $w=0$ mode from the Casimir energy. The $w\neq 0$ modes express winding modes and provide finite contributions to the $L$-dependence of the Casimir energy. On the other hand, $w=0$ {corresponds to} an unwinding mode and {it causes a} UV-divergence. Since the regularized Casimir energy $E^{\rm U(1)}_{\rm part2}[L]_{\rm reg.}$ does not contain any unwinding mode, it has no UV-divergence and {becomes} finite.} The explicit form of $E^{\rm U(1)}_{\rm part2}[L]$ is 
	\begin{align}
	E^{\rm U(1)}_{\rm part2}[L]_{\rm reg.}&=-\frac{3L}{32\pi^{5/2}}\sum^{\infty}_{w=1}\int^{\infty}_{0}dt\, t^{-\frac{7}{2}}e^{-\frac{w^2 L^2}{t}}\nonumber\\
	&=-\frac{3L}{32\pi^{5/2}}\sum^{\infty}_{w=1}\frac{1}{w^5 L^5}\int^{\infty}_{0}dt'\ t'{}^{\frac{5}{2}-1}e^{-t'}\nonumber\\
	&=-\frac{9}{128\pi^2 L^4}\zeta (5),
	\end{align}
where we performed the integration by substitution $t' \equiv \frac{w^2 L^2}{t}$ and used $\Gamma (\frac{5}{2}) =\frac{3\sqrt{\pi}}{4}$. From the above analysis, we obtain the regularized Casimir energy $E^{\rm U(1)}[L]_{\rm reg.}$ of the gauge field:
	\begin{align}
	E^{\rm U(1)}[L]_{\rm reg.}=-\frac{9}{128 \pi^2 L^4}\zeta (5).\label{U(1)Casimir}
	\end{align}
	
\begin{figure}[h]
	\begin{center}
	\includegraphics[width=7cm]{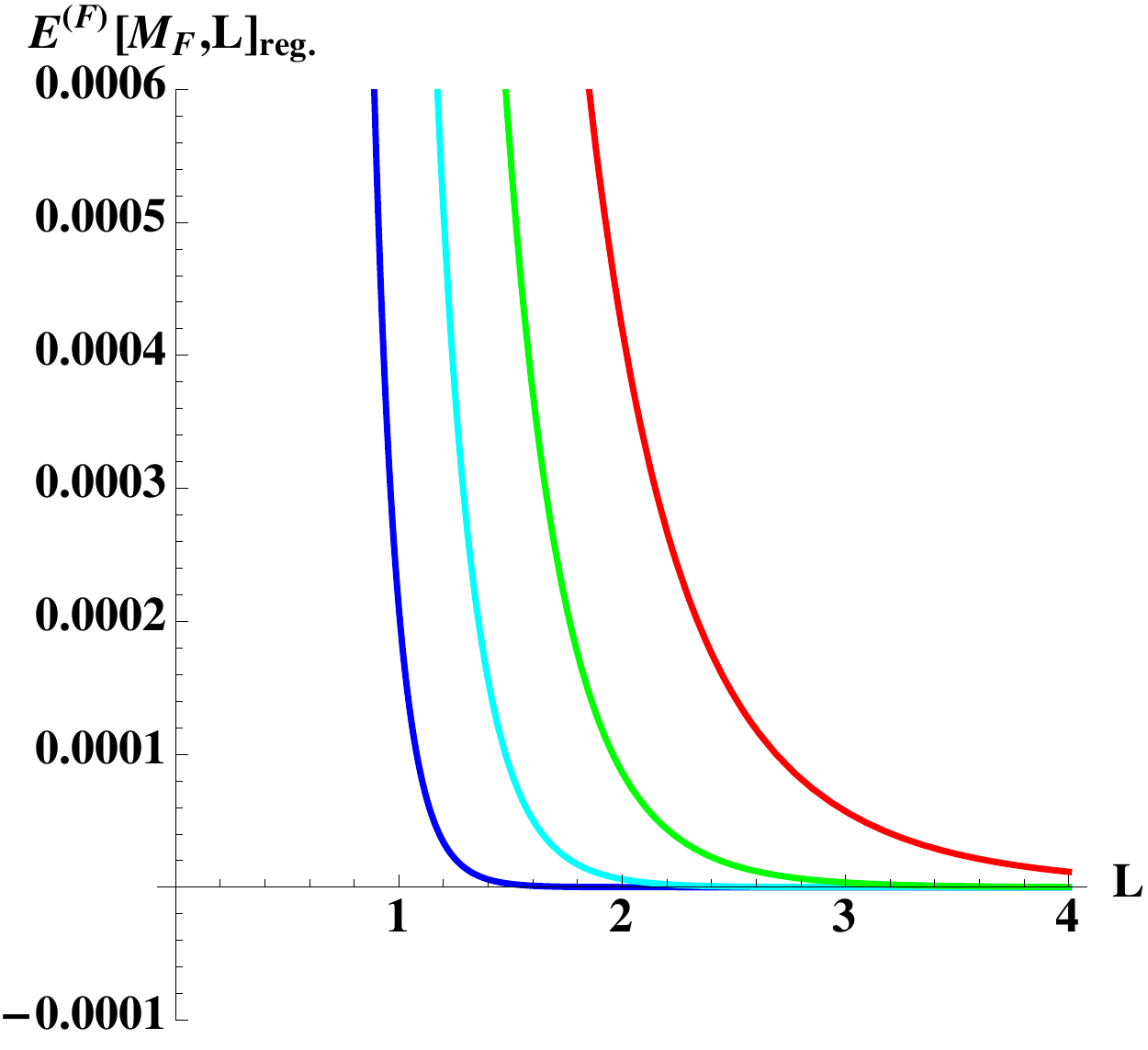}
	\caption{{Schematic} figure {of the $L$-dependence of the Casimir energy $E^{(F)}[M_F ,L]_{\rm reg.}$}. {The blue, cyan, green, {and red lines} correspond to the {cases} of $M_{F}=3.5$, $M_{F}=2$, $M_{F}=1.1$, {and} $M_{F}=0.4$, respectively}. {In this plot, $M_{F}$ and $L$ should be regarded as dimensionless parameters by multiplying a fundamental scale of the theory.}}
	\label{fig:E-fermion-0PI}
	\end{center}
	\end{figure}

\begin{figure}[h]
	\begin{center}
	\includegraphics[width=7cm]{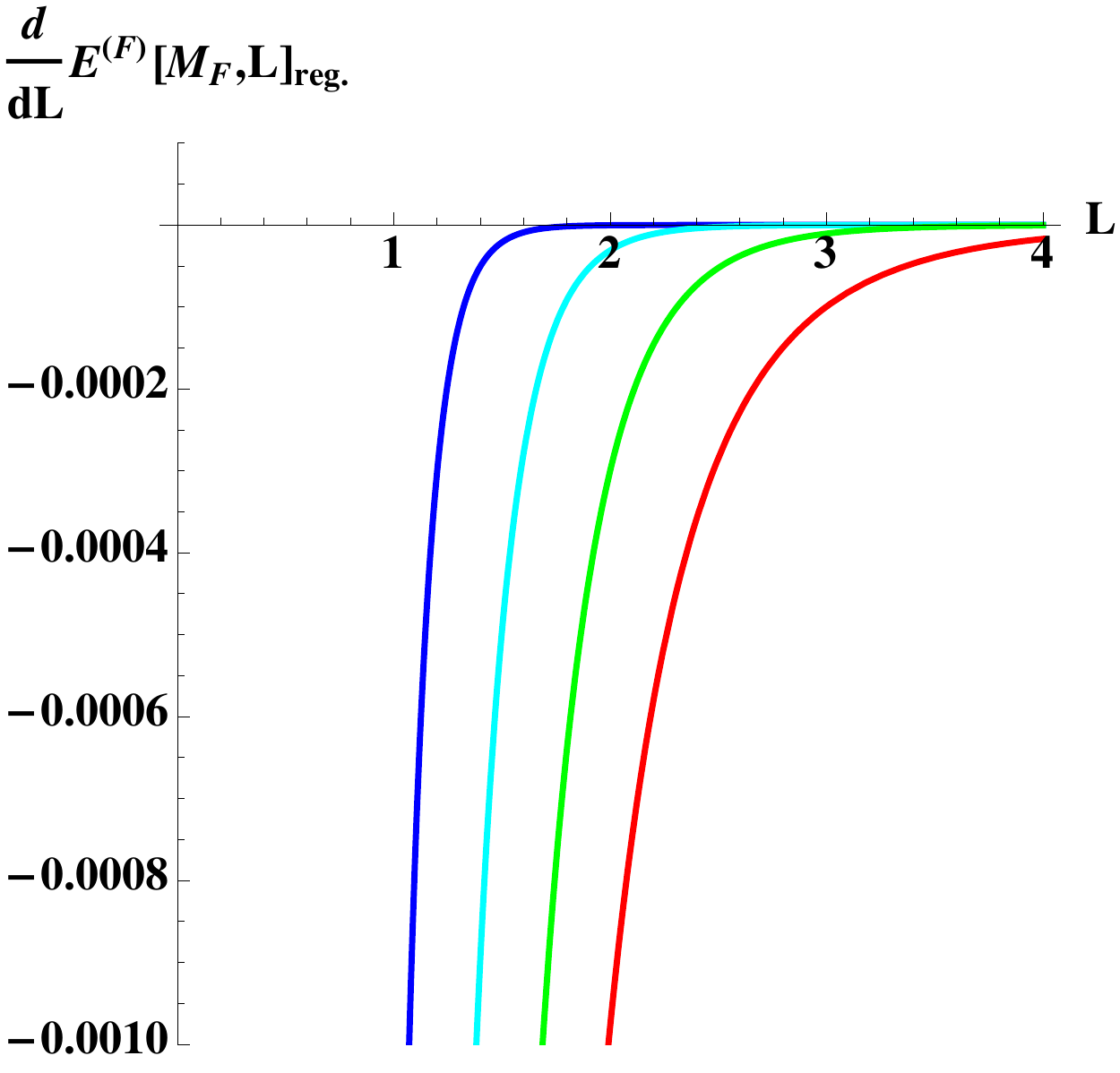}
	\caption{{Schematic} figure of {the $L$-dependence of the derivative of the Casimir energy $\frac{d}{dL}E^{(F)}[M_F ,L]_{\rm reg.}$}. {The blue, cyan, green, {and red lines} correspond to the {cases} of $M_{F}=3.5$, $M_{F}=2$, $M_{F}=1.1$, {and} $M_{F}=0.4$, respectively}. {In this plot, $M_{F}$ and $L$ should be regarded as dimensionless parameters by multiplying a fundamental scale of the theory.}}
	\label{fig:dEdL-fermion-0PI}
	\end{center}
	\end{figure}

{In the same way}, {we next} evaluate the Casimir energy of the fermion with the type-(II) boundary condition. {(}It is found that {the} type-(I) boundary condition {leads to} the same conclusion {as the type-(II)} for the Casimir energy.{)} To move on, we introduce the chiral representation:
	\begin{align}
	\psi^{(n)} =\left(\begin{array}{c}
				\xi^{(n)}\\
				0
				\end{array}
				\right)+\left(\begin{array}{c}
						0\\
						\eta^{(n)}
						\end{array}\right){.}
	\end{align}
The Gamma matrices are represented by 
	\begin{align}
	\gamma^{\mu}=\left(\begin{array}{cc}
		0&\bar{\sigma}^\mu\\
		\sigma^\mu &0\\
					\end{array}\right),
	\end{align}
where 
	\begin{align}
	\bar{\sigma}^\mu &=(1,{\bm \sigma}),\\
	\sigma^{\mu}&=(1,-{\bm \sigma}){,}
	\end{align}
{and} ${\bm \sigma}$ are Pauli matrices.
The partition function of the fermion {reads}
	\begin{align}
	Z_F [M_F ,L]&=\int [{d\Psi d\bar{\Psi}}]\ e^{iS_F}\nonumber\\
	&\propto {\rm exp}\left[ i\int d^4 x \Biggl\{ -i \int\frac{d^4 p}{(2\pi)^4}\, \biggl({\rm ln}\,p^\mu p_\mu +2\sum^{\infty}_{n=1}{\rm ln} \,(p^\mu p_\mu +m_{\psi^{(n)}}^2)\biggr)\Biggr\}\right]{,}
	\end{align}
{where the overall minus sign originates in the Grassmann property of fermions.}
After moving to the Euclidian space, we obtain the Casimir energy of the fermion:
	\begin{align}
	Z^{\rm Euclid}_F [M_F ,L]&\propto {\rm exp}\left[ -E^{(F)}[M_F ,L]\int d^4 x_{E}\right],
	\end{align}
where 
	\begin{align}
	E^{(F)}[M_F ,L]&= -\int \frac{d^4 p_E}{(2\pi)^4}\left[ {\rm ln}\, p_E^2 +2\sum^{\infty}_{n=1}\, {\rm ln}(p_E^2 +m_{\psi^{(n)}}^2)\right]\nonumber\\
	&=-\int \frac{d^4 p_E}{(2\pi)^4}\left[ {\rm ln}\,p_E^2 -{\rm ln}\, (p_E^2 +M_F^2)+\sum^{\infty}_{n=-\infty}{\rm ln}\, \biggl\{ p_E^2 +\Bigl(\frac{n\pi}{L}\Bigr)^2 +M_F^2\biggr\}\right].
	\end{align}
We divide $E^{(F)}[M_F ,L]$ into two parts as is the case of the gauge field:
	\begin{align}
	E^{(F)}[M_F ,L]&=E^{(F)}_{\rm part1}[M_F ,L]+E^{(F)}_{\rm part2}[M_F ,L],\\
	E^{(F)}_{\rm part1}[M_F ,L]&=-\int \frac{d^4 p_E}{(2\pi)^4}\left[ {\rm ln}\, p_E^2 -{\rm ln}\, (p_E^2 +M_F^2)\right],\\
	E^{(F)}_{\rm part2}[M_F ,L]&=-\int \frac{d^4 p_E}{(2\pi)^4}\sum^{\infty}_{n=-\infty}{\rm ln}\,\left[ p_E^2 +\Bigl(\frac{n\pi}{L}\Bigr)^2 +M_F^2\right].
	\end{align}
{In the same way as the gauge {field}, $E^{(F)}_{\rm part1}[M_F ,L]$ does not contain any $L$-dependence. Since we have an interest in the $L$-dependence of the Casimir energy, we just ignore this part.}
$E^{(F)}_{\rm part2}[M_F ,L]$ can {be also} evaluated as the gauge field case. {Using} the formulas (\ref{formula1})-(\ref{formula3}), we can rewrite $E^{(F)}_{\rm part2}[M_F ,L]$ as
	\begin{align}
	E^{(F)}_{\rm part2}[M_F ,L]=\frac{1}{16\pi^2}\sum^{\infty}_{n=-\infty}\int^{\infty}_{0}dt\, t^{-3}e^{-\left\{ (\frac{n\pi}{L})^2+M_F^2\right\}t}.
	\end{align}
By using the Poisson summation formula (\ref{poisson}), we obtain the following form for $E^{(F)}_{\rm part2}[M_F ,L]$:
	\begin{align}
	E^{(F)}_{\rm part2}[M_F ,L]=\frac{L}{16\pi^{5/2}}\sum^{\infty}_{w=-\infty}\int^{\infty}_{0}dt\,t^{-\frac{7}{2}}e^{-\frac{w^2 L^2}{t}-M_F^2 t}.
	\end{align}
Since $E^{(F)}_{\rm part2}[M_F ,L]$ contains UV-divergence {when} $t\rightarrow 0$, we regularize it as
	\begin{align}
	\frac{1}{L}E^{(F)}_{\rm part2}[M_F ,L]_{\rm reg.}\equiv \frac{1}{L}E^{(F)}_{\rm part2}[M_F ,L]-\frac{1}{L}E^{(F)}_{\rm part2}[M_F ,L]\Biggr|_{L\rightarrow \infty}.
	\end{align}
The regularized Casimir energy $E^{(F)}_{\rm part2}[M_F ,L]_{\rm reg.}$ is expressed by the modified Bessel function $K_\nu (z)$ as
	\begin{align}
	E^{(F)}_{\rm part2}[M_F ,L]_{\rm reg.}=\frac{{L}}{4\pi^{5/2}}\sum^{\infty}_{w=1} \left(\frac{|M_F |}{wL}\right)^{\frac{5}{2}}K_{\frac{5}{2}}(2w |M_F|L),
	\end{align}
where the modified Bessel function is defined by
	\begin{align}
	2\left(\frac{A}{B}\right)^{\frac{\nu}{2}} K_\nu (2\sqrt{AB})=\int^{\infty}_{0}dt\, t^{-\nu-1}e^{-At -\frac{B}{t}}. 
	\end{align}
Moreover, the modified Bessel function $K_{\frac{D}{2}}(z)$ with $D=\text{odd integer}$ can be expressed as 
	\begin{align}
	K_{\frac{D}{2}}(z)=\sqrt{\frac{\pi}{2z}}e^{-z}\sum^{\frac{D-1}{2}}_{k=0}\frac{\Bigl(\frac{D-1}{2}+k\Bigr)!}{k! \left(\frac{D-1}{2}-k\right)! (2z)^k}.
	\end{align}
Therefore the explicit form of $E^{(F)}_{\rm part2}[M_F ,L]_{\rm reg.}$ is given by
	\begin{align}
	E^{(F)}_{\rm part2}[M_F ,L]_{\rm reg.}=\frac{|M_F|^2}{8\pi^2 L^2}\sum^{\infty}_{w=1}\frac{e^{-2w|M_F|L}}{w^3}\left(1+\frac{3}{2w|M_F|L}+\frac{3}{4w^2 (|M_F|L)^2}\right).
	\end{align}
From the analysis, we obtain {the $L$-dependence of} the regularized total Casimir energy $E^{(F)}[M_F ,L]_{\rm reg.}$ of the fermion as
	\begin{align}
	{E^{(F)}[M_F ,L]_{\rm reg.}=\frac{|M_F|^2}{8\pi^2 L^2}\sum^{\infty}_{w=1}\frac{e^{-2w|M_F|L}}{w^3}\left(1+\frac{3}{2w|M_F|L}+\frac{3}{4w^2 (|M_F|L)^2}\right).\label{fermion-Casimir}}
	\end{align}
{Schematic figures} of the regularized Casimir energy of the fermion $E^{(F)}[M_F ,L]_{\rm reg.}$ and its derivative $\frac{d}{dL}E^{(F)}[M_F ,L]_{\rm reg.}$ are depicted in {Figure~\ref{fig:E-fermion-0PI} and Figure~\ref{fig:dEdL-fermion-0PI}}.
	
{Combining} all the results and concentrating on the {$L$-dependence} of the Casimir energy, we obtain the regularized Casimir energy $E[M_F ,L]_{\rm reg.}$ as
	\begin{align}
	E[M_F ,L]_{\rm reg.}&=E^{\rm U(1)}[L]_{\rm reg.}+E^{(F)}[M_F ,L]_{\rm reg.}\nonumber\\
	&=-\frac{9}{128 \pi^2 L^4}\zeta (5) {+\frac{|M_F|^2}{8\pi^2 L^2}\sum^{\infty}_{w=1}\frac{e^{-2w|M_F|L}}{w^3}\left(1+\frac{3}{2w|M_F|L}+\frac{3}{4w^2 (|M_F|L)^2}\right)}.\label{Casimir-U(1)model}
	\end{align}
{Schematic figures} of the total Casimir energy $E[M_F ,L]_{\rm reg.}$ and its derivative $\frac{d}{dL}E[M_F ,L]_{\rm reg.}$ are depicted in Figure~\ref{fig.E} and Figure~\ref{fig.dEdL}. We can find that {there} exists a non-trivial global minimum to the Casimir energy. Thus we can conclude that the extra dimension is stable in this setup. 

{We should give a comment} for the above results. {It was discussed in ref.~\cite{Ponton:2001hq} that,} in the case of $M_{F}=0$, {the} {$L$-dependence} of $E^{(F)}[M_F ,L]_{\rm reg.}$ becomes 
	\begin{align}
	E^{(F)}[M_F ,L]_{\rm reg.}\sim \dfrac{\alpha}{L^{4}}\quad (\alpha :\text{Const.}),
	\end{align}
	so that the {finite} global minimum does not appear in the Casimir energy. {In the case of $M_{F}\neq 0$, the fermion's positive contribution to the Casimir energy becomes dominant for $L\rightarrow 0$ because the fermion has more {degrees} of freedom than the gauge {field}. On the other hand, the negative contribution of the gauge {field} becomes dominant for $L\rightarrow \infty$ since the contribution of the fermion is suppressed by the exponential factor via the bulk mass. 
{Therefore, we have {revisited} that the extra dimension can be stabilized if the following two conditions, which were pointed out in Ref.~\cite{Ponton:2001hq}, are satisfied: (i) 5d massless gauge bosons exist and all 5d fermions have nonzero bulk masses{; and (ii) the }degrees of freedom of fermions are sufficiently larger than those of bosons.}
In our interval extra dimension case, in contrast with orbifold models, a bulk mass $M_{F}$ {is not forbidden} from any symmetry and should be involved so that the {finite} global minimum of the Casimir energy {can emerge}.
	\begin{figure}[H]
	\begin{center}
	\includegraphics[width=7cm]{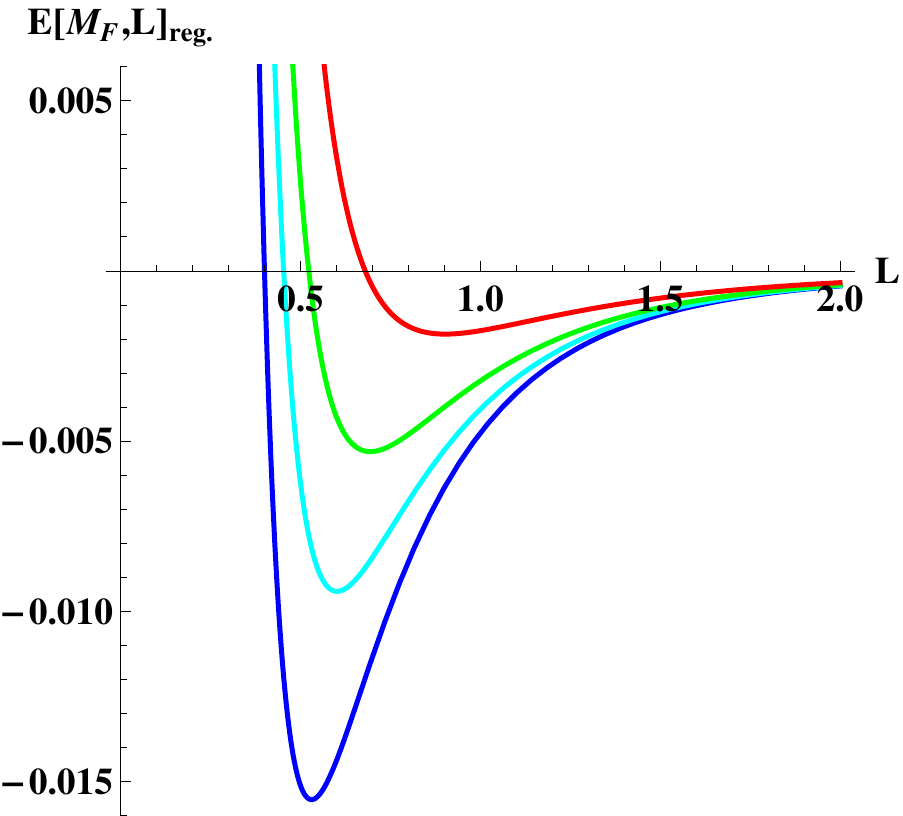}
	\end{center}
	\begin{center}
	\caption{{Schematic} figure of the total Casimir energy $E[M_F ,L]_{\rm reg.}$ as a function of the {length} $L$ of the extra dimension. The blue, cyan, green, {and red lines} correspond to the {cases} of {$M_F =1.7$, $M_F =1.5$, $M_F =1.3$, {and} $M_F =1$, respectively. In this plot, $M_{F}$ and $L$ should be regarded as dimensionless parameters by multiplying a fundamental scale of the theory.} We can find a non-trivial global minimum and can conclude that the extra dimension is stable.}
	\label{fig.E}
	\end{center}
	\end{figure}
	\begin{figure}[H]
	\begin{center}
	\includegraphics[width=7cm]{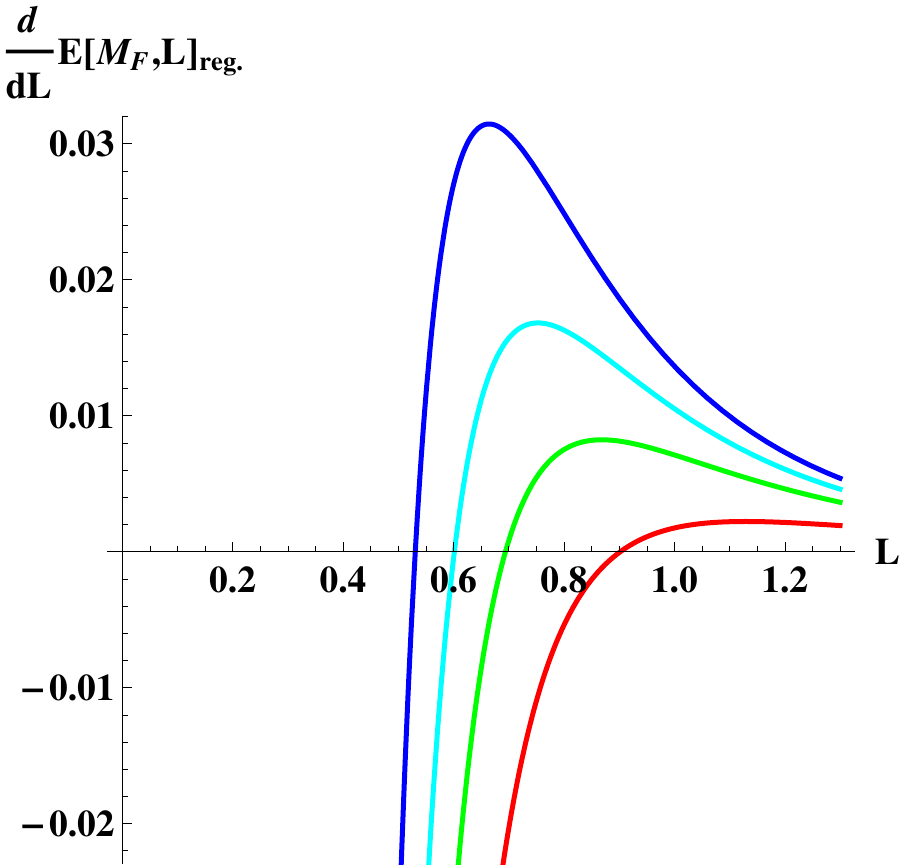}
	\caption{{Schematic} figure of the derivative of the total Casimir energy $\frac{d}{dL}E[M_F ,L]_{\rm reg.}$ as a function of the {length} $L$ of the extra dimension. The blue, cyan, green, {and red lines} correspond to the case of {$M_F =1.7$, $M_F =1.5$, $M_F =1.3${, and} $M_F =1$, respectively. In this plot, $M_{F}$ and $L$ should be regarded as dimensionless parameters by multiplying a fundamental scale of the theory}. We can find a non-trivial global minimum.}
	\label{fig.dEdL}
	\end{center}
	\end{figure}

\section{Theory with point interactions}\label{sec:Theory with point interactions}
In the papers \cite{Fujimoto:2012wv,Fujimoto:2013ki,Fujimoto:2014fka,Cai:2015jla}, a new way to produce generations and a mass hierarchy was proposed with introducing zero-width {branes}, {so-called point interactions,} to the extra dimension. In this section, we {briefly} review a theory with point interactions at first.
In the theory, massless zero modes {become} degenerate and {a nontrivial number of} generations {appears from a} {one-generation} 5d fermion {(where a self-contained comprehensive review on the formulation is provided in Appendix~\ref{sec:appendix})}.
{In this section, we clarify} the 4d mass spectrum of the theory {with point interactions}, which {plays} an important role in the calculation of the Casimir energy.
\subsection{BCs and 4d {mass} spectrum}
In a theory with point interactions, we can recognize the point {interactions} as {extra boundary points} and {need to} impose extra boundary conditions at the {points}. {Assuming} that only the fermion feels {the} {point interactions} at {$y=L_1,\, L_{2}$}, we can obtain {three-generation} chiral massless zero modes from the {{following BCs} :}
	\begin{align}
	\Psi_R (y)&=0\qquad \text{at}\qquad {y=0,\, L_1 {\pm \varepsilon},\, L_{2}{\pm \varepsilon},\, L},\label{RBC}\\
	&\text{or}\nonumber\\
	\Psi_L (y)&=0\qquad \text{at}\qquad {y=0,\, L_1 {\pm \varepsilon},\, L_{2}{\pm \varepsilon},\, L},\label{LBC}
	\end{align}
{where $\varepsilon$ represents an infinitesimal positive constant.} We should emphasize that the above BCs are consistent with the 5d gauge invariance since {they} are invariant under the 5d gauge transformation:
	\begin{align}
	\Psi_R (x,y)\ \rightarrow\ \widetilde{\Psi}_R (x,y)=e^{-ig\Lambda (x,y)}\Psi_R (x,y),\\
	\Psi_L (x,y)\ \rightarrow\ \widetilde{\Psi}_L (x,y)=e^{-ig\Lambda (x,y)}\Psi_L (x,y).
	\end{align}
We expand a 5d fermion $\Psi(x,y)$ {with the {BCs (\ref{RBC}) or (\ref{LBC})}:}
	\begin{align}
	{\Psi (x,y) =\Psi_R (x,y) +\Psi_L (x,y)=\sum_n \psi^{(n)}_R (x) \mathscr{F}^{(n)}_{\psi_R} (y)+\sum_n \psi^{(n)}_L (x)\mathscr{G}^{(n)}_{\psi_L} (y).}
	\end{align}
It was found in Refs.~\cite{Fujimoto:2012wv,Fujimoto:2013ki,Fujimoto:2014fka} that we have {three} {degenerate} zero modes {$\mathscr{G}^{(0)}_{i,\psi_L}(y)$} with {$i=1,2,3$ {$\bigl(\mathscr{F}^{(0)}_{i,\psi_R}$} with $i=1,2,3 \bigr)$} under the BC (\ref{RBC}) {$\bigl($the BC (\ref{LBC})$\bigr)$} and can obtain {three} {degenerate} massless chiral fermions {$\psi_{i,L}^{(0)}(x)$} {{$\bigl(\psi_{i,R}^{(0)}(x)\bigr)$}}:
	\begin{align}
	\Psi (x,y)&=\Psi_0 (x,y) +\sum_{n=1}^{\infty} \sum^{{3}}_{i=1}\left\{{\psi^{(n)}_{i,R}(x)} \mathscr{F}^{(n)}_{i,\psi_R} (y) +{\psi^{(n)}_{i,L}(x)} \mathscr{G}_{i,\psi_L}^{(n)} (y) \right\}, \label{ModeexpansionPIs}\\
	\Psi_0 (x,y)&= \left\{ \begin{array}{l}
					\displaystyle \sum^{{3}}_{i=1}{\psi^{(0)}_{i,L}(x)}\mathscr{G}_{i,\psi_L}^{(0)}(y),\qquad \text{{for}}\quad\Psi_R (y)=0\quad \text{at}\quad {y=0,\, L_1 {\pm \varepsilon},\, L_{2}{\pm \varepsilon},\, L},\\
					\displaystyle \sum^{{3}}_{i=1}{\psi^{(n)}_{i,R}(x)}\mathscr{F}_{i,\psi_R}^{(0)}(y),\qquad \text{{for}}\quad\Psi_L (y)=0\quad \text{at}\quad {y=0,\, L_1 {\pm \varepsilon},\, L_{2}{\pm \varepsilon},\, L},
				\end{array}\right.
	\end{align}
where $\mathscr{G}_{i,\psi_L}^{(0)} (y)$ {$\bigl(\mathscr{F}_{i,\psi_R}^{(0)}(y)\bigr)$} is a solution of eq.~(\ref{ZEROsolutionG}) {$\bigl($eq.~(\ref{ZEROsolutionF})$\bigr)$} under the BC~{(\ref{RBC}) with eq.~(\ref{BCRtoL})} {$\bigl($the BC{~(\ref{LBC}) with eq.~(\ref{BCLtoR})$\bigr)$}}. The explicit {forms} of $\mathscr{G}_{i,\psi_L}^{(0)}(y)$ {and $\mathscr{F}_{i,\psi_R}^{(0)}(y)$ are} {given by}
	\begin{align}
	\mathscr{G}_{i,\psi_L}^{(0)}(y)&=\sqrt{\frac{2M_F}{e^{2M_F l_i}-1}}e^{M_F (y-L_{i-1})} \Bigl[ \theta (y-L_{i-1})\theta(L_i-y)\Bigr]\\
	\mathscr{F}_{i,\psi_R}^{(0)}(y)&=\sqrt{\frac{2M_F}{1-e^{-2M_F l_i}}}e^{-M_F (y-L_{i-1})} \Bigl[ \theta (y-L_{i-1})\theta(L_i-y)\Bigr]
	\end{align}
where 
	\begin{align}
	{l_i \equiv L_i -L_{i-1}\quad (i=1,2,3;\ L_3 =L, L_0=0 )},\label{segment}
	\end{align}
and $\theta (y)$ is the step function. {Schematic figures} of the localized zero modes $\mathscr{G}_{i,\psi_L}^{(0)}(y)$ and $\mathscr{F}_{i,\psi_R}^{(0)}(y)$ {are} depicted in {Figure}~\ref{fig.G_0^i} and {Figure}~\ref{fig.F_0^i}. Each zero mode only lives in a segment and localizes to {a boundary.}

	\begin{figure}[h]
	\begin{center}
	\includegraphics[width=10cm]{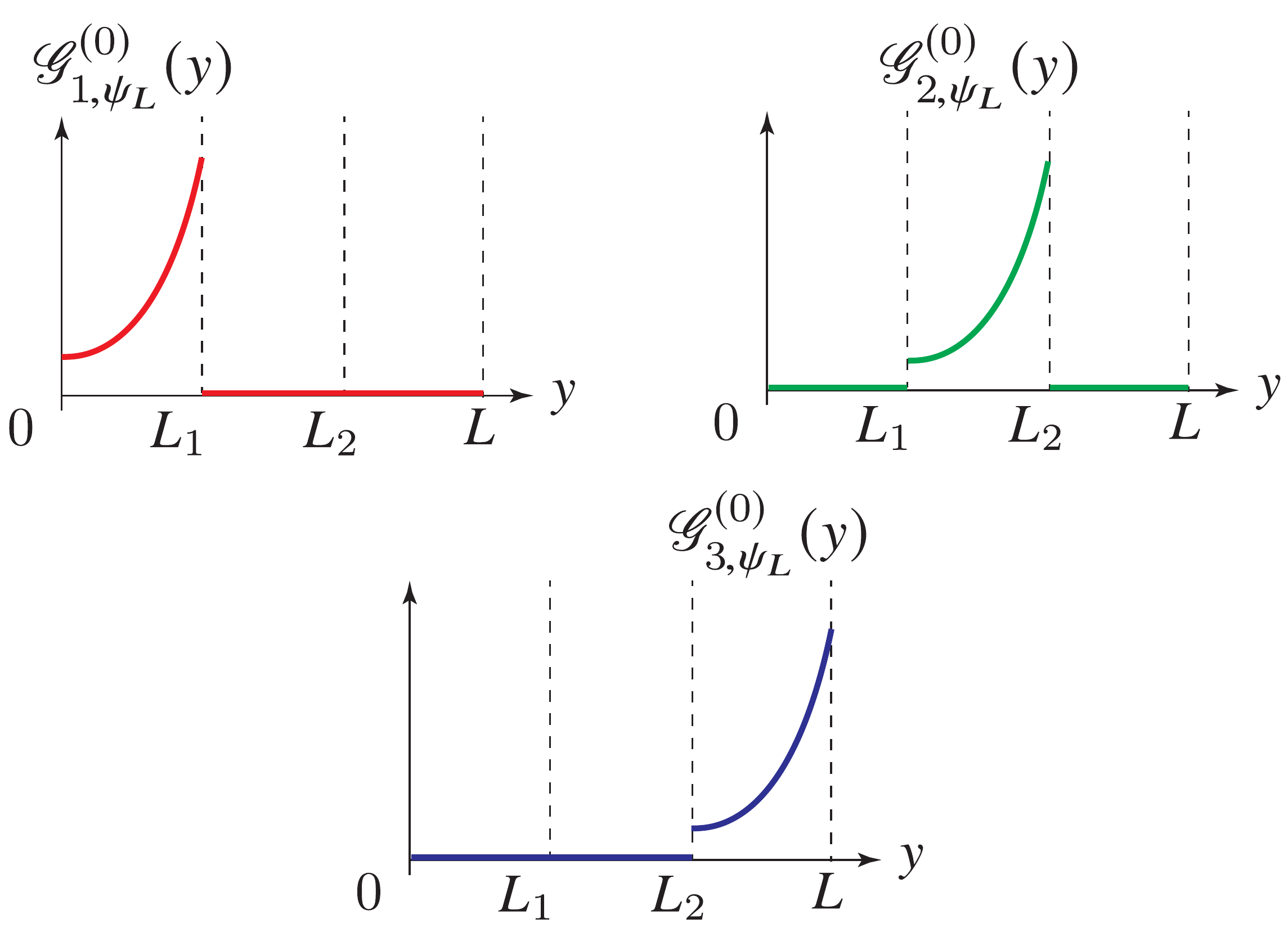}
	\vspace{-0.5cm}
	\caption{{Schematic figures} of localized zero modes {$\mathscr{G}^{(0)}_{i,\psi_L} (y)$ {($i=1,2,3$)} with $M_F >0$}. Each zero mode {of $\mathscr{G}^{(0)}_{i,\psi_L} (y)$} only has a {non-vanishing} value within {the} segment {$L_{i-1}<y<L_{i}$ and localizes to a boundary.}}
	\label{fig.G_0^i}
	\end{center}
	\end{figure}

	\begin{figure}[h]
	\begin{center}
	\includegraphics[width=10cm]{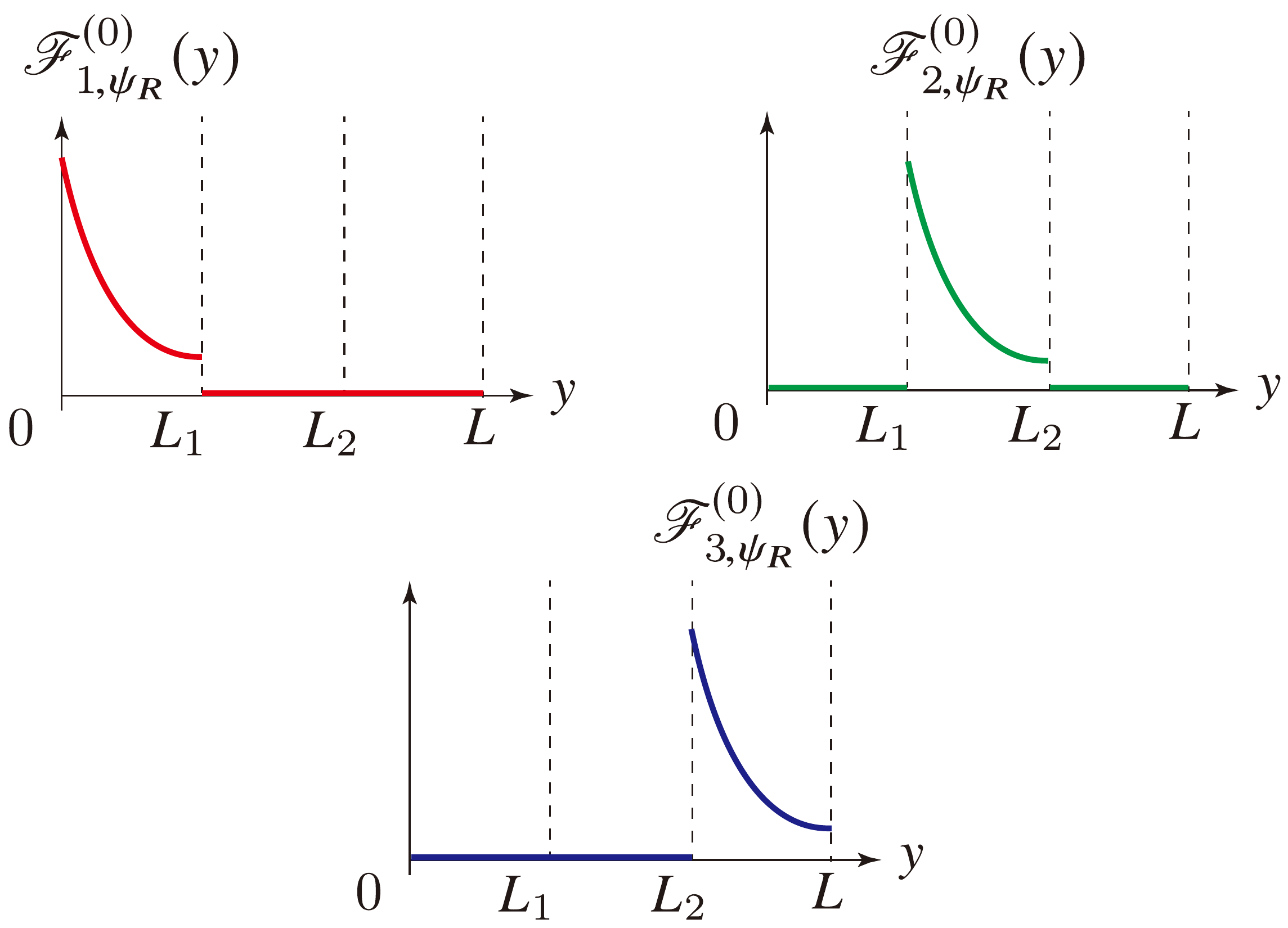}
	\vspace{-0.5cm}
	\caption{{Schematic figures} of localized zero modes {$\mathscr{F}^{(0)}_{i,\psi_R} (y)$ {($i=1,2,3$)} with $M_F >0$}. Each zero mode {of $\mathscr{F}^{(0)}_{i,\psi_R} (y)$} only has a {non-vanishing} value within {the} segment {$L_{i-1}<y<L_{i}$ and localizes to a boundary.}}
	\label{fig.F_0^i}
	\end{center}
	\end{figure}

After substituting eq.~(\ref{ModeexpansionPIs}) into the action (\ref{Fermion_action}) and using the {orthonormal} {relations}
	\begin{align}
	\int^{L}_{0}dy\,{\bigl(\mathscr{F}^{(n)}_{i,\psi_R}(y)\bigr)^{\ast}} \mathscr{F}^{(m)}_{j,\psi_R} (y)&=\delta_{n,m}\delta_{i,j},\\
	\int^{L}_{0}dy\,{\bigl(\mathscr{G}^{(n)}_{i,\psi_L} (y)\bigr)^{\ast}}\mathscr{G}^{(m)}_{j,\psi_L} (y)&=\delta_{n,m}\delta_{i,j},\quad {(i,j=1,2,3)}
	\end{align}
we obtain {the} 4d spectrum of the fermion{,}
	\begin{align}
	S_F =\int d^4 x \left\{ {\cal L}_{n=0} +\sum_{n=1}^{\infty}{\sum^{3}_{i=1}}\ \bar{\psi}^{(n)}_{i} (i\gamma^{\mu}\partial_{\mu} +m_{i,\psi^{(n)}} )\psi^{(n)}_{i}\right\},
	\end{align}
where
	\begin{align}
	{\cal L}_{n=0}=\left\{\begin{array}{l}
					\displaystyle {\sum^{3}_{i=1}}\ \psi^{(0)}_{i,L}{(x)}(i\gamma^\mu \partial_\mu )\psi^{(0)}_{i,L}{(x)} \qquad \text{for the {BC (\ref{RBC})}},\\
					\displaystyle {\sum^{3}_{i=1}}\ \psi^{(0)}_{i,R}{(x)}(i\gamma^\mu \partial_\mu )\psi^{(0)}_{i,R}{(x)} \qquad \text{for the {BC (\ref{LBC})}}.\end{array}\right.
	\end{align}
{and the} 4d mass spectrum $m_{i,\psi^{(n)}}$ is given by
	\begin{align}
	m_{i,\psi^{(n)}}={\sqrt{M_F^2 +\left(\frac{n\pi}{l_i}\right)^2}}\qquad {({i=1,2,3;}\ n=1,2,3,\cdots)},
	\end{align}
where $l_i$ is defined by eq.~(\ref{segment}).

\section{{Dynamical generation of fermion mass hierarchy}}\label{Semi-realistic}
In this section, by using the previous results, {we consider an $SU(2)\times U(1)$ model with a single generation of 5d fermions, which produces three generations of 4d chiral fermions by the point interactions, and discuss whether the model can dynamically generate a fermion mass hierarchy.} {To this end, we first set an action and BCs of {this} model. {The action consists of an $SU(2)$ gauge field, a $U(1)$ gauge field, a single generation $SU(2)$ doublet fermion, a single generation $SU(2)$ singlet fermion{,} and an $SU(2)$ doublet scalar field.  {The contents of our model {mimic} those of the SM without the color degree of freedom{, where the $U(1)$ (hyper)charges of $Q$ and $U$ take those of the quark doublet and the up-type singlet.}}} Extra BCs via point interactions are {a} key ingredient to produce the {three} generations from one generation 5d fermion as we reviewed in Section \ref{sec:Theory with point interactions}{. The} positions of the point interactions {crucially} {affect} the fermion mass hierarchy through the overlap integrals, {as} we will see in Section \ref{subsec:Mass hierarchy}. {We will show that the positions of the point interactions can} be determined dynamically through the minimization of the Casimir energy {and then find that an exponential fermion mass hierarchy naturally {appears}}. Following the results, {we discuss the stability of the extra dimension.}

{\subsection{Action and BCs}}\label{sec:Action and BCs_Semi-realistic}
\noindent {{We start with} the following action {for the gauge fields and fermions}:}
	\begin{align}
	S&=S_G +S_F,\\
	S_G&=\int d^4 x \int^L_0 dy \left[-\frac{1}{4}W^{aMN}W_{MN}^{a}-\frac{1}{2}(\partial^M W_M^a)^2 -i\,\bar{c}^a (\partial^M {\cal D}_M)c^a\right.\nonumber\\
	& \hspace{10em}\left.{-\frac{1}{4}F^{MN}F_{MN}-\frac{1}{2}(\partial^{M}A_{M})^2 -i\bar{c}(\partial^{M}\partial_{M})c}\right],\\
	S_F&=\int d^4 x\int^L_0 dy \left[\bar{Q}\Bigl(i\Gamma^M D_M^{(Q)}+M_F^{(Q)}\Bigr)Q+{\bar{U}\Bigl(i\Gamma^M \partial_M +M_F^{(U)}\Bigr)U}\right],\label{5d-fermions-action}
	\end{align}
where
	\begin{align}
	W^a_{MN}&=\partial_M W_N^a -\partial_N W_M^a {-g\varepsilon_{abc}W_M^b W_N^c},\\
	{F_{MN}}&{=\partial_{M}A_{N}-\partial_{N}A_{M},}\\
	{\cal D}_M c^a &=\partial_M c^a +g\varepsilon_{abc}W^b_M c^c,\\
	D_M^{(Q)} Q&=\biggl(\partial_M {-igW_M^a T_a{-ig'A_{M}}}\biggr)Q.
	\end{align}
{$W_M^{a}$, $A_{M}$, {$c^{a}$}, $c$ and {$\bar{c}^{a}$}, $\bar{c}$ denote {an} $SU(2)$ gauge, a $U(1)$ gauge, {ghost and anti-ghost fields}, respectively.} {$g$ and $g'$ denote $SU(2)$ and $U(1)$ couplings of the $SU(2)$ doublet fermion.} $Q$ and {$U$} indicate {an} $SU(2)$ doublet fermion and {an} $SU(2)$ single fermion{, respectively}. A bulk mass of the 5d fermion is denoted by $M_F^{(\Psi )}$ ($\Psi =Q,{U}$). $\varepsilon_{abc}$ is {a} complete antisymmetric tensor and $T_a$ is a generator of $SU(2)$ {acts on a fundamental representation}, which satisfies the following algebra and the orthogonal relation:
	\begin{align}
	[T_a, T_b]=i\varepsilon_{abc}T_c,\\
	\text{tr}\,T_a T_b=\frac{1}{2}\delta_{a,b}.
	\end{align}
{According to the analysis given in Section \ref{sec:4d spectum of a 5d $U(1)$ gauge theory on an interval}, we} choose boundary conditions for the fields as follows:
	\begin{align}
	&\left\{
		\begin{array}{l}
		\partial_y W_\mu^{a}(x,y)=0,\\
		W_y^{a}(x,y)=0,
		\end{array}
	\right.
	\hspace{2em}\text{at}\qquad y=0,L,\label{NABC}\\
	&\left\{
		\begin{array}{l}
		{\partial_y c^{a}(x,y)=0},\\
		{\partial_y \bar{c}^{a}(x,y)=0},
		\end{array}
	\right.
	\hspace{3em}\text{at}\qquad y=0,L,\\[0.2cm]
	&{\left\{
		\begin{array}{l}
		\partial_y A_\mu(x,y)=0,\\
		A_y(x,y)=0,
		\end{array}
	\right.
	\hspace{2em}\text{at}\qquad y=0,L,}\label{ABC}\\
	&{\left\{
		\begin{array}{l}
		\partial_y c(x,y)=0,\\
		\partial_y \bar{c}(x,y)=0,
		\end{array}
	\right.
	\hspace{3em}\text{at}\qquad y=0,L,}\\[0.2cm]
	&Q_R (x,y)=0\hspace{5em} \text{at}\qquad y=0,L_1 {\pm \varepsilon},L_2 {\pm \varepsilon} ,L,\\
	&{U_L  (x,y)}=0\hspace{5em}\text{at}\qquad y=0,L_1 {\pm \varepsilon},L_2 {\pm \varepsilon} ,L{,}
	\end{align}
{where $L_{1}$ and $L_{2}$ ($0<L_{1}<L_{2}<L$) denote the positions of the point interactions and {$\varepsilon$ represents an infinitesimal positive constant.}}
A schematic figure of the extra dimension is depicted in {Figure}~\ref{fig.semi-realistic_extra_dim}. We {introduced} two point interactions {at $y=L_{1}$ and $L_{2}$} for the fermions and put the situation that all fermions feel the point interactions {at} the same positions for {simplicity}. {On the other hand, the gauge and the ghost fields are assumed not to feel the point interactions at $y=L_{1}$ and $L_{2}$. {We note that the 5d gauge symmetries are intact under the configuration of the boundary conditions.}}
	\begin{figure}[h]
		\begin{center}
		\includegraphics[width=9cm]{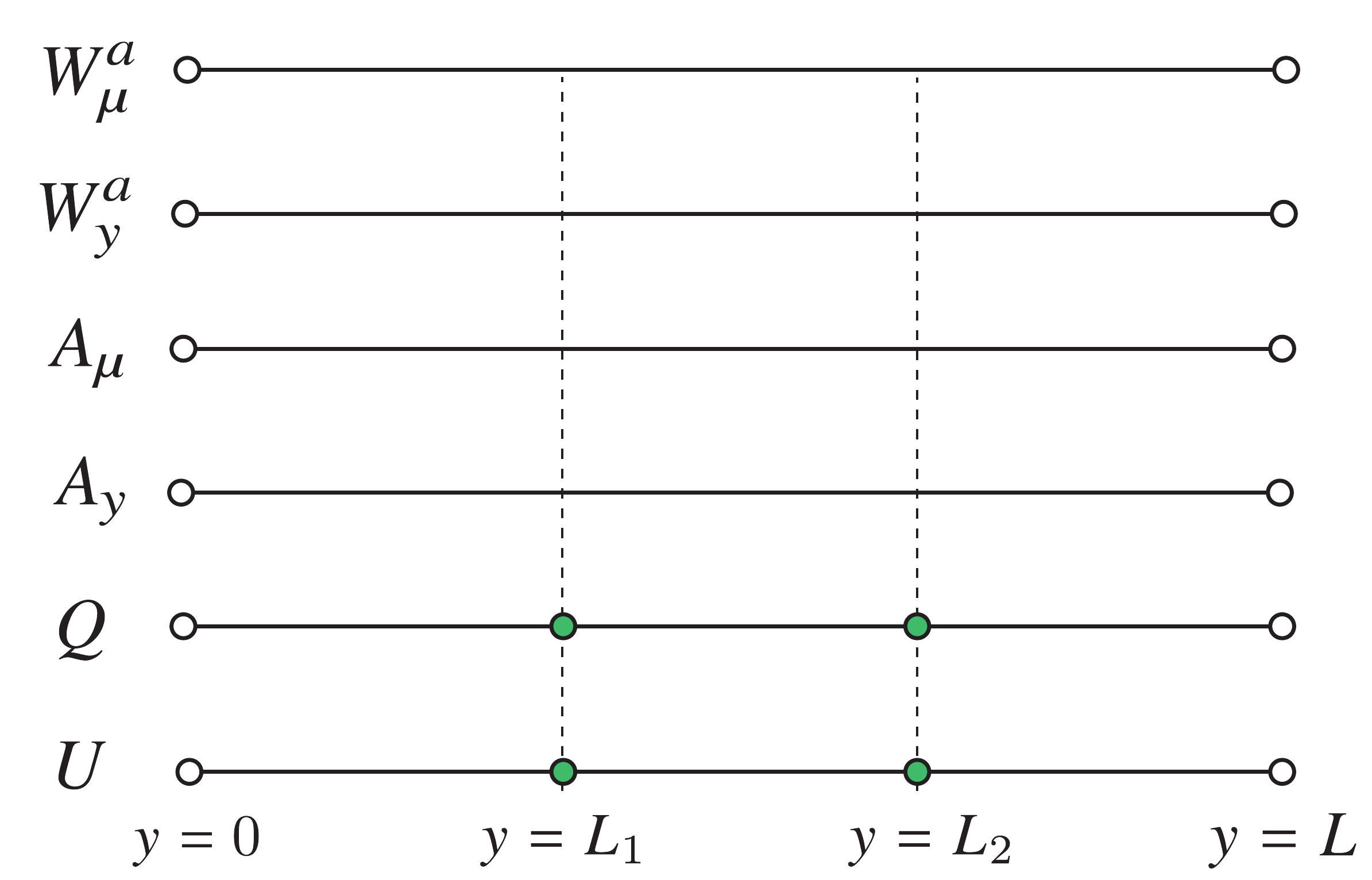}
		\caption{A schematic figure of the extra dimension. Only the fermions $Q$ and $U$ feel the point interactions (Green dots) at $y=L_1,L_2$ and {the gauge fields $W_\mu^{a}$, $W_y^{a}$, $A_{\mu}$ and $A_{y}$} do not. The situation is completely consistent with 5d gauge invariance.}
		 \label{fig.semi-realistic_extra_dim}
		 \end{center}
	\end{figure}

{\subsection{Determination of the positions of the point interactions}\label{sec:Determination of the positions of the point interactions}}
\noindent {Using the results of section~\ref{sec.2-3}, we can evaluate the Casimir energy as a function of the positions of {the} point interactions $\{L_{1}, L_{2}\}$}:
	\begin{align}
	&{E^{{(F)}}[M^{(Q)}_{F},M^{{(U)}}_{F},L_{1},L_{2},L]_{\rm reg.}}\nonumber\\
	{=}&\,2\cdot\frac{|M^{(Q)}_{F}|^2}{8\pi^2 L_{1}^{2}}\sum^{\infty}_{w=1}\frac{{\rm e}^{-2w|M^{(Q)}_{F}|L_{1}}}{w^3}\left(1+\frac{3}{2w|M_{F}^{(Q)}|L_{1}}+\frac{3}{4w^2 |M_{F}^{(Q)}|^2 L_{1}^2}\right)\nonumber\\
	&+2\cdot \frac{|M_{F}^{(Q)}|^2}{8\pi^2 (L_{2}-L_{1})^2}\sum^{\infty}_{w=1}\frac{{\rm e}^{-2w|M_{F}^{(Q)}|(L_{2}-L_{1})}}{w^3}\left(1+\frac{3}{2w|M_{F}^{(Q)}|(L_{2}-L_{1})}+\frac{3}{4w^2 |M_{F}^{(Q)}|^2(L_{2}-L_{1})^2}\right)\nonumber\\
	&+2\cdot \frac{|M_{F}^{(Q)}|^2}{8\pi^2 (L-L_{2})^2}\sum^{\infty}_{w=1}\frac{{\rm e}^{-2w|M_{F}^{(Q)}|(L-L_{2})}}{w^3}\left(1+\frac{3}{2w|M_{F}^{(Q)}|(L-L_{2})}+\frac{3}{4w^2 |M_{F}^{(Q)}|^2(L-L_{2})^2}\right)\nonumber\\
	&+\frac{|M^{{(U)}}_{F}|^2}{8\pi^2 L_{1}^{2}}\sum^{\infty}_{w=1}\frac{{\rm e}^{-2w|M^{{(U)}}_{F}|L_{1}}}{w^3}\left(1+\frac{3}{2w|M^{{(U)}}_{F}|L_{1}}+\frac{3}{4w^2 {|M^{{(U)}}_{F}|}^2 L_{1}^2}\right)\nonumber\\
	&+\frac{|M^{{(U)}}_{F}|^2}{8\pi^2 (L_{2}-L_{1})^2}\sum^{\infty}_{w=1}\frac{{\rm e}^{-2w|M^{{(U)}}_{F}|(L_{2}-L_{1})}}{w^3}\left(1+\frac{3}{2w|M^{{(U)}}_{F}|(L_{2}-L_{1})}+\frac{3}{4w^2 |M^{{(U)}}_{F}|^2(L_{2}-L_{1})^2}\right)\nonumber\\
	&+\frac{|M^{{(U)}}_{F}|^2}{8\pi^2 (L-L_{2})^2}\sum^{\infty}_{w=1}\frac{{\rm e}^{-2w|M^{{(U)}}_{F}|(L-L_{2})}}{w^3}\left(1+\frac{3}{2w|M^{{(U)}}_{F}|(L-L_{2})}+\frac{3}{4w^2|M^{{(U)}}_{F}|^2(L-L_{2})^2}\right). \label{Vefffreefermion-2PI}
	\end{align}
{{With the fixed {length} $L$,} the minimization condition for the Casimir energy can determine the {values} of the parameters $\{L_{1},L_{2}\}$. The above potential {turns out to have} the {finite} global minimum at $L_{1}=\frac{L}{3}$, $L_{2}=\frac{2L}{3}$. To verify this statement, we consider the following function $I(x,y,z)$:
	\begin{align}
	&I(x,y,z)=f(x)+f(y)+f(z),\label{imitatefunction}\\
	&{x,y,z> 0},\\
	&x+y+z=1\label{constraintcondition}.
	\end{align}
$I(x,y,z)$ {imitates} the function form of the fermion Casimir energy with the variables $x=\widetilde{L}_{1}$, $y=\widetilde{L}_{2}-\widetilde{L}_{1}$, $z=1-\widetilde{L}_{2}$, where $\widetilde{L}_{i}$ ($i=1,2$) is defined as $\widetilde{L}_{i}\equiv L_{i}/L$. We assume the function $f(x)$ {to be} a {monotonically} decreasing function {and also {$f' (x)\equiv \dfrac{df(x)}{dx}$} to be} a {monotonically} increasing one {with} $\displaystyle{\lim_{x\to 0}}f(x)=+\infty$. {We note that the fermion Casimir energy~(\ref{fermion-Casimir}) turns out to satisfy those assumptions (see Figures~\ref{fig:E-fermion-0PI} and \ref{fig:dEdL-fermion-0PI}).} Substituting the condition eq.~(\ref{constraintcondition}) into eq.~(\ref{imitatefunction}), we obtain
	\begin{align}
	I(x,y,1-x-y)=f(x)+f(y)+f(1-x-y).
	\end{align}
To investigate an extreme value of the above function, we examine $\dfrac{\partial I}{\partial x}$ and $\dfrac{\partial I}{\partial y}$: 
	\begin{align}
	\frac{\partial I}{\partial x}&=f'(x)-f'(1-x-y),\\
	\frac{\partial I}{\partial y}&=f'(y)-f'(1-x-y),
	\end{align}
From the conditions $\dfrac{\partial I}{\partial x}=0$ and $\dfrac{\partial I}{\partial y}=0$, {we obtain the result}
	\begin{align}
	f'(x)=f'(y)=f'(1-x-y){.}\label{f-condition}
	\end{align}
Since we assumed that {$f'(x)$} is a {monotonically} increasing function, the result (\ref{f-condition}) {can be realized only when}
	\begin{align}
	x=y=z=\frac{1}{3}.
	\end{align}
Thus we find that $I(x,y,z)$ has {an} extreme value when $x=y=z=\dfrac{1}{3}$. Moreover, the function takes a local minimum {at $x=y=z=\dfrac{1}{3}$}. {To show this}, we consider the second-order differentials with the condition $x=y=z=\dfrac{1}{3}$:
	\begin{align}
	&\frac{\partial^{2} I}{\partial x^{2}}\biggr|_{x=y=\frac{1}{3}}=2f''\Bigl(\frac{1}{3}\Bigr),\\
	&\frac{\partial^{2} I}{\partial y\partial x}\biggr|_{x=y=\frac{1}{3}}=f''\Bigl(\frac{1}{3}\Bigr),\\	
	&\frac{\partial^{2} I}{\partial x\partial y}\biggr|_{x=y=\frac{1}{3}}=f''\Bigl(\frac{1}{3}\Bigr),\\	
	&\frac{\partial^{2} I}{\partial y^{2}}\biggr|_{x=y=\frac{1}{3}}=2f''\Bigl(\frac{1}{3}\Bigr).
	\end{align}
We now consider the Hessian matrix $M$:
	\begin{align}
	M=
	\left(
	\begin{array}{cc}
	\dfrac{\partial^2 I}{\partial x^2}\biggl|_{x=y=\frac{1}{3}}& \dfrac{\partial^2 I}{\partial x\partial y}\biggl|_{x=y=\frac{1}{3}}\\[0.2cm]
	\dfrac{\partial^2 I}{\partial y\partial x}\biggl|_{x=y=\frac{1}{3}}&\dfrac{\partial^2 I}{\partial y^2}\biggl|_{x=y=\frac{1}{3}}
	\end{array}
	\right)=\left(\begin{array}{cc}
			2f''\Bigl(\frac{1}{3}\Bigr)& f''\Bigl(\frac{1}{3}\Bigr)\\
			f''\Bigl(\frac{1}{3}\Bigr)&2f''\Bigl(\frac{1}{3}\Bigr)
			\end{array}\right).
	\end{align}
Since {$f'(x)$} is a {monotonically} increasing function, $f''(x)>0$. Thus we find that 
	\begin{align}
	{\rm tr}\,M>0,\\
	{\rm det}\, M>0.
	\end{align}
The above results imply that the eigenvalues of the matrix $M$ are positive {and hence that} the position $x=y=\dfrac{1}{3}$ is a local minimum of the potential. Moreover, there is no other stationary point, we found that the position $x=y=\dfrac{1}{3}$ is a global minimum of the function $I(x,y,z)$.
From the above {discussions}, we conclude that the Casimir energy (\ref{Vefffreefermion-2PI}) has a global minimum at {$L_{1}=\frac{L}{3}$, $L_{2}=\frac{2L}{3}$.}}

\subsection{{Fermion mass hierarchy}}\label{subsec:Mass hierarchy}
Under the {above} situation, we can produce the fermion mass hierarchy dynamically {by} introducing the Yukawa coupling {to} {an $SU(2)$ doublet scalar field $\Phi (x,y)$}, which possesses the y-dependent VEV \footnote{{The $SU(2)$ doublet scalar may be regarded as $i\sigma_{2}H^{\ast}$ ($H$ is {the Higgs field}) {in the standard notation}.}}
	\begin{align}
	{\langle\Phi(y)\rangle}  &{=\left(\begin{array}{c}
									\phi(y)\\
									0
									\end{array}\right),}\label{S-VEV}\\
	{\phi(y)}&{={\cal A}e^{My}},
	\end{align}
as in eq.~(\ref{y-dependentVEV}) because of the Robin boundary condition (\ref{Robin_BCs}).\footnote{
{As we discussed in~\cite{Fujimoto:2012wv}, if the warped scalar
VEV is provided in the Higgs doublet,
a serious violation in the gauge universality is expected. Thereby,
in the previous works~\cite{Fujimoto:2012wv,Fujimoto:2013ki,Fujimoto:2014fka}, we introduced an additional singlet scalar with
the warped VEV, while the Higgs doublet has the ordinary constant VEV, 
and considered 'higher-dimensional' Yukawa terms where {both} the Higgs
doublet and the singlet scalar appear.
(Note that we prohibited the ordinary Yukawa terms by introducing the $Z_2$
 discrete symmetry: odd parity for the two scalars, even parity for the others.)
In this manuscript, we assumed that the Higgs doublet contains the warped VEV
for displaying the form of the Yukawa terms in a simple way, for avoiding {confusion}
originating from why the two types of the scalars are introduced.
This 'simplification' is just making the explanation simplified, and
mass matrix takes basically the same form between in this 'simplified' setup and in the original
setup without gauge universality violation.
See the related discussion in section~\ref{sec:Conclusion and Discussion}.}
}	 	 
The {situation,} in which the {$i$-th generation ($i=1,2,3$)} fermion lives in the segment $y\in [L_{i-1},L_{i}]$ ($L_{0}\equiv0, L_{3}\equiv L$) and the {scalar field} lives in every {region,} makes a large mass hierarchy for the fermion masses through the Yukawa interaction {$\lambda \bar{Q}\Phi {U}$}:
	\begin{align}
	m_{i}= \lambda \int^{L}_{0}dy \, {\bigl(\mathscr{G}^{(0)}_{i,Q_{L}}(y)\bigr)^{\ast}}{\phi(y)} \mathscr{F}^{(0)}_{i,{{U_{R}}}}(y){,}\qquad {(i=1,2,3)}
	\end{align}
A schematic figure is depicted in Figure~\ref{fig.mass-hierarchy}. Since the minimization of the Casimir energy {determines} the positions of the point interactions as to make the {distances} between them equal, the exponential VEV of the {scalar field} makes an exponential mass hierarchy {such as }
	\begin{align}
	{\frac{m_{2}}{m_{1}}=\frac{m_{3}}{m_{2}}={e^{\frac{1}{3}ML}}.}
	\end{align}
{Thus, the fermion mass hierarchy around $10^{5}$ can be obtained by suitably choosing the parameter {$ML$}.}
We emphasize that this mass hierarchy appears dynamically since the positions of the point interactions and the form of the VEV of the {scalar} {are determined} dynamically.
	\begin{figure}[h]
		\begin{center}
		\includegraphics[width=9cm]{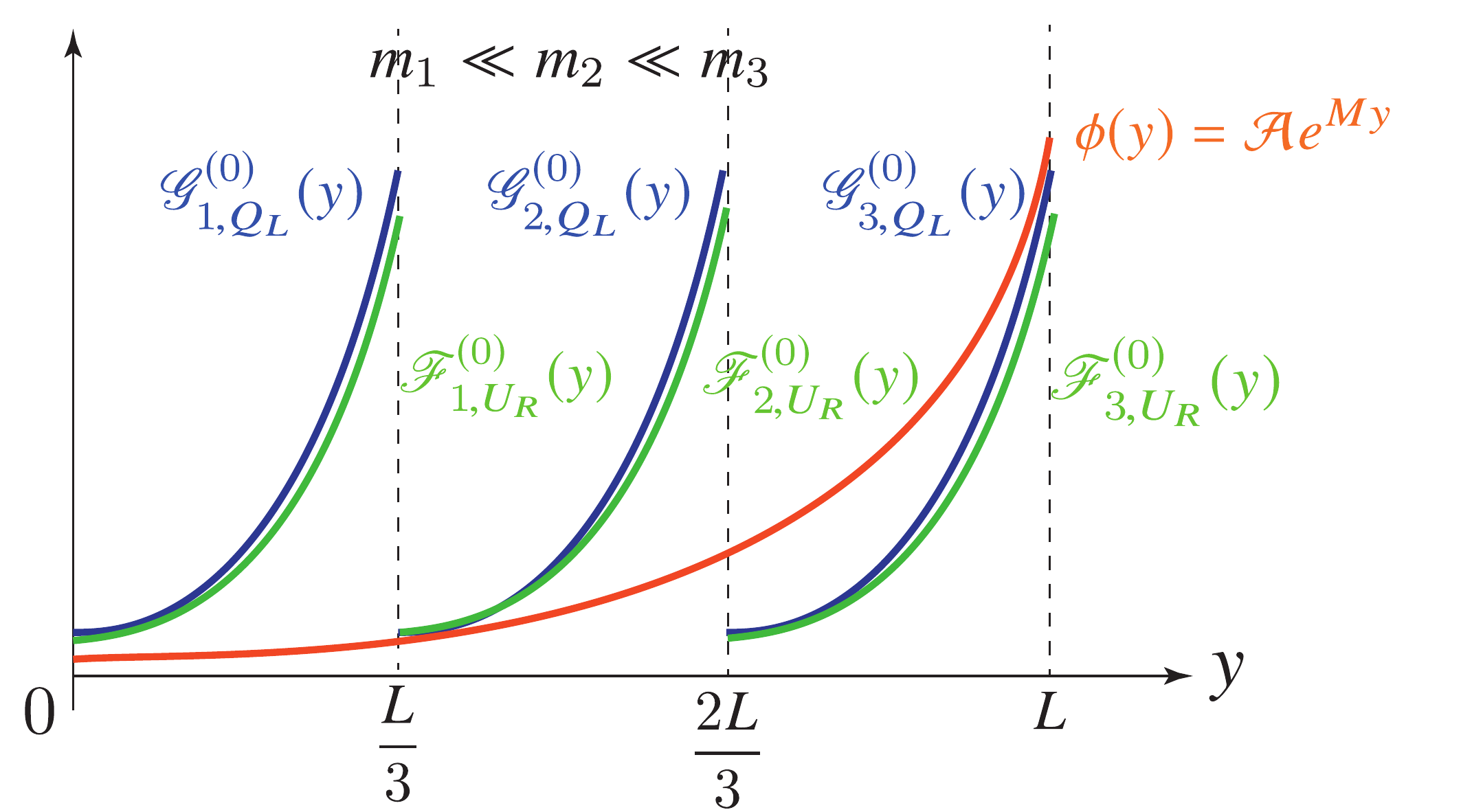}
		\caption{{Schematic} figure of {zero mode profiles of chiral massless fermions and the VEV of {the scalar field $\phi(y)$}}. The figure is depicted with the situation $M_{F}^{(Q)}>0$ and {$M_{F}^{{(U)}}<0$}. The position of the point interactions are fixed by the minimization of the Casimir energy and the $y$-dependent {scalar} VEV {produces an exponential} mass hierarchy through the overlap integrals with respect to the extra dimension.}
		 \label{fig.mass-hierarchy}
		 \end{center}
	\end{figure}

{\subsection{Stability of the extra dimension}}
\noindent {We have shown that for} any {fixed} {length} $L$, the positions of the point interactions are determined dynamically to the value $L_{1}=\frac{L}{3}$, $L_{2}=\frac{2L}{3}$ from the minimization of the Casimir energy. Under this situation, {we discuss the stability of the {whole} extra dimension. {In} our model the $SU(2)$ doublet scalar field $\Phi(x,y)$ possesses the y-dependent VEV and breaks the gauge symmetry as $SU(2)\times U(1)\to U(1)'$. Therefore, we will discuss the stability of the extra dimension in the {broken phase.}}

As we investigated in Section 3, the extra dimension can be stabilized if the following two conditions are satisfied: (i) 5d massless gauge bosons exist and all 5d fermions have nonzero bulk masses. (ii) The degrees of freedom of fermions are sufficiently larger than those of bosons. The first condition (i) will ensure that the Casimir energy approaches to zero with negative values in $L \to \infty$ limit, as in (3.42). The second condition (ii) will ensure that the Casimir energy goes to $+\infty$ in $L \to 0$ limit, as in (3.42).

{In our model, the $SU(2) \times U(1)$ gauge symmetry is broken by the VEV of the scalar but a subgroup $U(1)'$ is still unbroken. Thus, the first condition (i) is satisfied in our model. The second condition (ii) seems to be satisfied in our model because the degrees of freedom of the fermions become three times the number of 5d fermions due to the point interactions. Moreover, there is still room for introducing extra fermions by using {the} }type-(III) {BCs}, which do not produce any exotic chiral massless fermions. Therefore, in our setup, the extra dimension is expected to be {stabilized} by the Casimir energy.\footnote{In the full SM-like setup, the gluon and the scalar contribute to the Casimir energy. To determine the value of the length $L$ of the extra dimension, we need to calculate the Casimir energy of all the fields in the gauge symmetry broken phase with the $y$-dependent VEV, which is beyond the scope of this paper. }

\section{Conclusion and Discussion}\label{sec:Conclusion and Discussion}
In this paper, we {proposed} a new mechanism to produce a fermion mass hierarchy dynamically by introducing the point interactions to the 5d gauge theory on an interval. The interval extra dimension can possibly {be} stable and the point interactions produce generations {of fermions}. The positions of the point interactions were determined by minimizing the Casimir energy of the {fermions. The} extra-dimension coordinate-dependent VEV of the {scalar field}, which is also produced dynamically {under the Robin boundary condition}, {makes exponentially different} fermion masses through the overlap integrals.

{We give a comment for the contribution of the {scalar field} to the Casimir energy at first. In this paper, we ignored the effect of the {scalar field} to the Casimir energy {for simplicity} {because the contribution {to} the Casimir energy from the {scalar field} will have no exact analytic expression due to the Robin BC. {However,} the inclusion of the {scalar field} {will not change} the conclusions about the stability of the {whole} extra dimension and the positions of the {point interactions if} the degrees of freedom of the fermions are sufficiently {larger than} those of bosons.}} 

Next, some comments are given to the flavor mixing of the fermions. In our model, we introduced the point interactions at $y=L_{1}, L_{2}$ for {both the} $SU(2)$ doublet and {the} singlet {fermions}. {Here, mass matrices are diagonal and flavor mixing cannot appear.} In general, however, there is no need to share the point interactions {in} fermions so that we can introduce the {individual} point interactions to {each fermion}, respectively{, which means that (e.g.)} the $SU(2)$ doublet fermion feels the point interactions at $y=L_{1},L_{2}$ and the $SU(2)$ singlet fermion feels the point interactions at $y=L'{}_{1},L'{}_{2}$ {\cite{Fujimoto:2012wv,Fujimoto:2013ki,Fujimoto:2014fka}}{. Then the} mode functions of {the} {$SU(2)$-doublet} zero mode $\mathscr{G}^{(0)}_{i,Q_{L}}(y)$ and {the} {$SU(2)$-singlet} zero mode $\mathscr{F}^{(0)}_{j,{{U_{R}}}}(y)$ may have an overlap {for $i\neq j$}. In other words, off diagonal components may appear {in} the mass matrix as
	\begin{align}
	m_{ij}=\lambda \int^{L}_{0}dy \, {\Bigl(\mathscr{G}^{(0)}_{i,Q_{L}}(y)\Bigr)^{\ast}}{\phi(y)} \mathscr{F}^{(0)}_{j,{{U_{R}}}}(y),\qquad {(i,j=1,2,3)}
	\end{align}
and a flavor mixing can be realized.

If the minimization of the Casimir energy determines the positions of the point interactions as $L_{1}=L'{}_{1}$, $L_{2}=L'{}_{2}$, flavor mixing does not appear so that we need an idea to make $L_{1}\neq L'{}_{1}$, $L_{2}\neq L'{}_{2}$. One way {to avoid the situation of $L_{i}=L'{}_{i}$} is to consider higher loop effects of the Casimir energy, which may make $L_{1}\neq L'{}_{1}$, $L_{2}\neq L'{}_{2}$ through the interactions. {To introduce exotic 5d fermions, {where they contribute to the Casimir energy,} is another way. No chiral massless zero modes appear {in an} exotic {fermion} {when we assign} a suitable choice of boundary conditions {to it.} {Under such conditions,} the low energy matter {contents of} the model, i.e. the {Standard Model} {particles, do} not have a change. If we put a different boundary condition to $y=L_{1}$ ($y=L'_{1}$) from $y=L_{2}$ ($y=L'_{2}$), each {segment} has a different contribution of the Casimir energy so that we may produce the flavor mixing.} {A different strategy} is to introduce more than two point interactions, e.g. $N-1$ point interactions for the $SU(2)$ doublet fermion and $N'-1$ point interactions for the $SU(2)$ singlet fermion, {where we} divide the interval extra dimension into more than three segments, i.e. $N$ segments for the doublet and $N'$ segments for the singlet. A combination of type-(I)$\bigl($type-(II)$\bigr)$ and type-(III) BCs {can} produce three massless zero modes for $SU(2)$ doublet and singlet fermions, respectively. In this situation, the minimization of the Casimir energy determines the positions of $N-1$ ($N'-1$) point interactions and zero modes of the $SU(2)$ doublet (singlet) appear {in} three of the $N$ ($N'$) segments. {A suitable choice of the segments with zero modes} may possibly produce {off-diagonal} components of the mass matrix, i.e. flavor mixing {even after taking account of the stabilization of the point interactions}. 

Finally, we {focus on} the gauge universality. It was pointed out in Refs.~\cite{Fujimoto:2012wv,Fujimoto:2013ki,Fujimoto:2014fka} that the gauge symmetry breaking due to the y-dependent VEV of the scalar field {would} cause {a} gauge universality violation. That is because the y-dependent VEV of the scalar modifies the flat profile of the zero mode function of the gauge boson and thereby the values of the {4d} gauge couplings change with respect to the generations through the overlap integrals. A way to avoid this crisis is to introduce two scalar fields; one is an $SU(2)$ doublet scalar and another is a gauge-singlet scalar field. In the situation that the constant VEV of the $SU(2)$ doublet scalar breaks the gauge symmetry and the y-dependent VEV of the gauge-singlet scalar provides a mass hierarchy, we can {avoid} the gauge universality violation. It would be of great interest to construct a more phenomenologically {viable} model along the lines discussed in this paper.

\section*{Acknowledgement}
We thank Nobuhito Maru for discussions in the early stage of this work. This work is supported in part by Grants-in-Aid for Scientific Research [No.~15K05055 and No.~25400260 (M.S.)] from the Ministry of Education, Culture, Sports, Science and Technology (MEXT) in Japan.

\appendix
\section*{Appendix}
\section{{Review of point interactions and derivation of fermion profiles
under the existence of one point interaction in the bulk of an interval}}
\label{sec:appendix}
In this Appendix, 
we first review a one-dimensional quantum {mechanical} system with a point interaction,
and then apply the formulation for the five-dimensional Dirac action on an interval
with a point interaction.

The well-known Dirac $\delta$-function potential in quantum mechanics is 
an example of the point interaction, i.e., the interaction of zero range.
The consistent manner to treat such a singularity has been given in Ref.~\cite{Cheon:2000tq}.
According to the formulation~\cite{Cheon:2000tq},
we regard a point interaction as an idealized long wavelength or infrared limit
of localized interactions in one dimension, and
hence it is a singular interaction with zero range at one point, say $y=L_{1}$
on a line $\mathbb{R}$.
A system with such an interaction can be described by the system on the line
with the singular point removed, namely, on $\mathbb{R}\backslash\{L_{1}\}$.
In order to construct a quantum system on the domain $D = \mathbb{R}\backslash\{L_{1}\}$,
we require that the probability current 
$j_{y}(y)=-i\bigl((\partial_{y}\varphi^{\ast})\varphi
-\varphi^{\ast}(\partial_{y}\varphi)\bigr)(y)$
is continuous around the singular point, i.e.~\cite{Cheon:2000tq}
%
\begin{align}
j_{y}(L_{1}-\varepsilon)=j_{y}(L_{1}+\varepsilon)\,,
\label{A.1}
\end{align}
%
where $\varepsilon$ represents an infinitesimal positive constant.
We note that the above probability conservation guarantees 
the Hemiticity of the Hamiltonian.

The requirement (\ref{A.1}) implies that any state in the domain $D$ must obey 
a certain set of {BCs} at $y = L_{1} \pm \varepsilon$.
For example, the Dirichlet BC 
%
\begin{align}
\varphi(L_{1}-\varepsilon) = 0 = \varphi(L_{1}+\varepsilon)
      \label{A.2}
\end{align}
%
satisfies the condition (\ref{A.1}).
The Dirichlet BC may be understood as a point interaction given by
the Dirac $\delta$-function potential $V(y)=\alpha\delta(y)$ 
with the limit of the coupling $\alpha \rightarrow \infty$.
Another type of boundary conditions, which also satisfies the condition (\ref{A.1}),
is known as the Robin BC, i.e.
%
\begin{align}
-\varphi^{\prime}(L_{1}-\varepsilon) + M \varphi(L_{1}-\varepsilon) 
 = 0 
 = -\varphi^{\prime}(L_{1}+\varepsilon) + M \varphi(L_{1}+\varepsilon)\,,
      \label{A.3}
\end{align}
%
where $\varphi^{\prime}(y)=\frac{\partial\varphi(y)}{\partial y}$
and $M$ is a constant parameter with mass-dimension one.
The above two types of BCs will become important in our analysis.

Interestingly, point interactions can appear not only in one-dimensional quantum mechanics 
but also in extra dimension scenarios \cite{Fujimoto:2012wv},
since finite-range-localized interactions will be regarded as the point interaction 
in an idealized long wavelength or infrared limit, 
like a domain wall potential or a brane in extra dimension models. 
Hence, as we will see below, we apply the above point interaction treatment 
to the five-dimensional Dirac action on an interval ($y \in [0, L]$) 
with a point interaction at $y=L_{1}$,
and explain how to decompose a five-dimensional
Dirac field $\Psi(x,y)$ into KK mass eigenmodes, in a self-contained way.

The five-dimensional free action for $\Psi(x,y)$ that we focus on is
%
\begin{align}
S = \int d^4 x \left[ \int_0^{{L_1-\varepsilon}} dy 
     + \int_{{L_1+\varepsilon}}^L dy  \right] 
        \overline{\Psi}(x,y) \left( i \partial_M \Gamma^M + M_F \right) \Psi(x,y)\,,
\label{eq_app:5d_action}
\end{align}
%
where the system contains an extra specific point at $y = L_1$ in addition to the two end points of the interval at $y = 0, L$, and is divided into two parts by the presence of the {point interaction} at $y = L_1$. {Using the knowledge of the point interaction on one-dimensional quantum mechanical systems, we }describe profiles of KK particles appearing in one-extra-dimensional scenarios.

We first need to find a consistency requirement like the probability current conservation
(\ref{A.1}) in quantum mechanics with a point interaction.
Such a consistency requirement of our system is a current conservation along 
the $y$-direction, i.e.\footnote{
The consistency requirement will be obtained from the action principle
$\delta S =0$~\cite{Csaki:2003dt}.
In this case, the derived conditions are given by
%
\begin{align}
\big[ \overline{\Psi} \Gamma_y \delta \Psi \big]_{y = 0} 
 = 0\
 = \big[ \overline{\Psi} \Gamma_y \delta \Psi \big]_{y = L}\,&, 
\label{eq_app:BC_form_1} \\
\big[ \overline{\Psi} \Gamma_y \delta \Psi \big]_{y = L_1 - \varepsilon} 
 = \big[ \overline{\Psi} \Gamma_y \delta \Psi \big]_{y = L_1 + \varepsilon}\,&,
\label{eq_app:BC_form_2}
\end{align}
%
where $\delta \Psi$ means the variation of $\Psi$.
Since $\Psi$ and $\delta\Psi$ can be regarded as independent fields with the
assumption that  $\Psi$ and $\delta \Psi$ take the same boundary conditions,
it seems that the above conditions are more restrictive than those of 
(\ref{current_conservation_1}) and (\ref{current_conservation_2}).
However, they turn out to reduce the same {conclusions} in our analysis given below,
{and in fact the Dirichlet BC (\ref{DirichetBC_Psi}) satisfies
(\ref{eq_app:BC_form_1}) and (\ref{eq_app:BC_form_2}).}
The above conditions have been analyzed in {Ref.~\cite{Csaki:2003sh} (see also \cite{Fujimoto:2011kf,Fujimoto:2016llj,Fujimoto:2016rbr})} and we will not discuss 
{(\ref{eq_app:BC_form_1}) and (\ref{eq_app:BC_form_2})} here.
}
%
\begin{align}
&j_{y}(x,y) \big|_{y=0} = 0 = j_{y}(x,y) \big|_{y=L}\,, 
\label{current_conservation_1}\\
&j_{y}(x,y) \big|_{y=L_{1}-\varepsilon} 
  = j_{y}(x,y) \big|_{y=L_{1}+\varepsilon}\,,
\label{current_conservation_2}
\end{align}
%
where
%
\begin{align}
j_{y}(x,y) \equiv \big(\overline{\Psi} \Gamma_y \Psi\big)(x,y)\,.
\label{current_def}
\end{align}
%
The conditions (\ref{current_conservation_1}) imply that 
there should be no current flow in the $y$-direction outside of the two ends of 
the interval.
The {condition} (\ref{current_conservation_2}) can be understood 
as the current conservation in the $y$-direction at the point interaction.

Since the current form $\overline{\Psi} \Gamma_y \Psi$ is equivalent to 
$\overline{\Psi_L} \Gamma_y \Psi_R + \overline{\Psi_R} \Gamma_y \Psi_L${,}
the Dirichlet BC 
%
\begin{align}
\Psi_R = 0 \quad \text{or} \quad \Psi_L = 0
 \quad  \textrm{at}\ y = 0, L_{1}-\varepsilon, L_{1}+\varepsilon, L
\label{DirichetBC_Psi}
\end{align}
%
is found to satisfy the conditions (\ref{current_conservation_1}) and 
(\ref{current_conservation_2}).
Thus, in the following analysis, we take the {BCs}
%
\begin{align}
\Psi_R = 0 \quad \text{at } y = 0, \ L_1 - \varepsilon, \ L_1 + \varepsilon , {L}\,.
\label{eq_app:BC_Psi_R_zero}
\end{align}
%
We should emphasize that once the {BCs} for the right-handed
part of $\Psi$ are fixed as above, the boundary conditions for the opposite
chirality, i.e. the left-handed part of $\Psi$ are automatically {determined}
through the equation of motion as\footnote{%
We cannot impose the Dirichlet BC for both 
$\Psi_{R}$ and $\Psi_{L}$ at a boundary because it is enough for
$\Psi_{R}=0$ or $\Psi_{L}=0$ to satisfy the conditions
(\ref{current_conservation_1}) and (\ref{current_conservation_2}),
and in fact the requirement $\Psi_{R} = \Psi_{L} =0$ at a boundary
is overconstrained.%
}
%
\begin{align}
(- \partial_y + M_F) \Psi_L = 0 \quad \text{at } y = 0, \ L_1 - \varepsilon, \ L_1 + \varepsilon , {L}\,.
\label{eq_app:associated_BC}
\end{align}
%
It is {worthwhile} noticing that wavefunctions $\Psi_{R/L}(y)$ and/or their derivatives 
$\Psi'_{R/L}(y)$ will become discontinuous at $y=L_{1}$, as we will see later,
because the continuity conditions of $\Psi_{R/L}(y=L_{1}-\varepsilon)=\Psi_{R/L}(y=L_{1}+\varepsilon)$ and $\Psi'_{R/L}(y=L_{1}-\varepsilon)=\Psi'_{R/L}(y=L_{1}+\varepsilon)$ are {{\it not}} imposed {on} $\Psi_{R/L}(y)$ here.

To perform the KK decomposition of the {5d} fields $\Psi_{L}(x,y)$ and $\Psi_{R}(x,y)$, let us consider the following one-dimensional eigenvalue equations:
	\begin{align}
	(-\partial_{y}^{2}+M_{F}^{2}){\mathscr{G}(y)} &=m^{2}{\mathscr{G}(y)}\quad \ \text{on } D\,,\label{eigen-1}\\
	(-\partial_{y}^{2}+M_{F}^{2}){\mathscr{F}(y)}  &=m^{2}{\mathscr{F}(y)}\quad \text{on } D\,,\label{eigen-2}
	\end{align}
with the BCs
	\begin{align}
	(-\partial_{y}+M_{F}){\mathscr{G}(y)}&=0\qquad \text{at}\quad y=0,L_{1}-\varepsilon, L_{1}+\varepsilon,L\,,\label{BCf}\\
	{\mathscr{F}(y)}&=0\qquad \text{at}\quad y=0,L_{1}-\varepsilon, L_{1}+\varepsilon,L\,,\label{BCg}
	\end{align}
where the one-dimensional domain $D$ is defined by
	\begin{align}
	&D=D_{1}\cup D_{2}\,,\\
	&\left\{\begin{array}{l}
	D_{1}=\{y\,|0\leq y<L_{1}\}\,,\\
	D_{2}=\{y\,|L_{1}<y\leq L\}\,.
	\end{array}\right.
	\end{align}
The eigenfunctions of the equations (\ref{eigen-1}) and (\ref{eigen-2}) with the BCs (\ref{BCf}) and (\ref{BCg}) are found to be of the form
%
	\begin{align}
	{\mathscr{G}^{(0)}_{1,\psi_{L}}(y)}&=\begin{cases}
	\sqrt{\frac{2M_{F}}{e^{2M_{F}{L_{1}}}-1}}e^{M_{F}{y}} \qquad &\text{on }D_{1},\\
	\hspace{1cm}0\qquad &\text{on }D_{2},
	\end{cases}
	\label{eq_app:specific_profile_first} \\
	{\mathscr{G}^{(n)}_{1,\psi_{L}}(y)}&=\begin{cases}
	{\frac{1}{m^{(n,1)}}\sqrt{\frac{2}{L_{1}}}}
	\left\{\frac{n\pi}{{L_{1}}}\cos\left(\frac{n\pi {y}}{{L_{1}}}\right)
	+M_{F}\sin\left(\frac{n\pi {y}}{{L_{1}}}\right)\right\}\qquad &\text{on }D_{1},\\
	\hspace{1cm}0\qquad &\text{on }D_{2},
	\end{cases}\\
	{\mathscr{F}^{(n)}_{1,\psi_{R}}(y)}&=\begin{cases}
	\sqrt{\frac{2}{{L_{1}}}}\sin\left(\frac{n\pi {y}}{{L_{1}}}\right)
	 \ \ \ \qquad &\text{on }D_{1},\\
	\hspace{1cm}0\qquad &\text{on }D_{2},	
	\end{cases}\\
	{\mathscr{G}^{(0)}_{2,\psi_{L}}(y)}&=\begin{cases}
	\hspace{1cm}0\qquad&\text{on }D_{1},\\
	 \sqrt{\frac{2M_{F}}{e^{2M_{F}{(L-L_{1})}}-1}}e^{M_{F}(y-L_{1})} 
	 \qquad &\text{on }D_{2},
	\end{cases} 
	\label{eq_app:specific_profile_first_2} \\
	{\mathscr{G}^{(n)}_{2,\psi_{L}}(y)}&=\begin{cases}
	\hspace{1cm}0\qquad &\text{on }D_{1},\\
	{\frac{1}{m^{(n,2)}}\sqrt{\frac{2}{L-L_{1}}}}
	\left\{\frac{n\pi}{{L-L_{1}}}\cos\left(\frac{n\pi (y-L_{1})}{{L-L_{1}}}\right)
	+M_{F}\sin\left(\frac{n\pi (y-L_{1})}{{L-L_{1}}}\right)\right\}\qquad &\text{on }D_{2},
	\end{cases}\\
	{\mathscr{F}^{(n)}_{2,\psi_{R}}(y)}&=\begin{cases}
	\hspace{1cm}0\ \qquad&\text{on }D_{1},\\
	\sqrt{\frac{2}{{L-L_{1}}}}\sin\left(\frac{n\pi (y-L_{1})}{{L-L_{1}}}\right)
	 \ \ \ \qquad&\text{on }D_{2},	
	\end{cases}\label{eq_app:specific_profile_last}
	\end{align}
%
{with $n=1,2,\cdots$.} We notice that even though the eigenfunctions {$\mathscr{G}^{(n)}_{1,\psi_{L}}(y)$ and $\mathscr{F}^{(n')}_{1,\psi_{R}}(y)$ $\bigl(\mathscr{G}^{(n)}_{2,\psi_{L}}(y)$ and $\mathscr{F}^{(n')}_{2,\psi_{R}}(y)\bigr)$} entirely vanish on $D_{2}$ (on $D_{1}$),
they are well defined on the whole domain $D$ and satisfy the eigenvalue equations
	\begin{align}
	(-\partial_{y}^{2}+M_{F}^{2}){\mathscr{G}^{(n)}_{i,\psi_{L}}(y)}=\bigl({m_{i,\psi^{(n)}}}\bigr)^{2}{\mathscr{G}^{(n)}_{i,\psi_{L}}(y)}\quad \text{on }D,\\
	(-\partial_{y}^{2}+M_{F}^{2}){\mathscr{F}^{(n')}_{i,\psi_{R}}(y)}=\bigl({m_{i,\psi^{(n')}}}\bigr)^{2}{\mathscr{F}^{(n')}_{i,\psi_{R}}(y)}\quad \text{on }D,
	\end{align}
and the BCs
	\begin{align}
	(-\partial_{y}+M_{F}){\mathscr{G}^{(n)}_{i,\psi_{L}}(y)}&=0\qquad \text{at}\quad y=0,L_{1}-\varepsilon, L_{1}+\varepsilon,L,\\
	{\mathscr{F}^{(n')}_{i,\psi_{R}}(y)}&=0\qquad \text{at}\quad y=0,L_{1}-\varepsilon, L_{1}+\varepsilon,L,
	\end{align}
for $n=0,1,2,\cdots$, $n'=1,2,\cdots$, $i=1,2$ with the eigenvalues
{
%
	\begin{align}
	{m_{1,\psi^{(0)}}}&=0,&(i=1,2)\,,\\
	{m_{1,\psi^{(n)}}}&=\sqrt{M_{F}^{2}+\Big(\dfrac{n\pi}{L_{1}}\Big)^{2}}\,,&(n=1,2,\cdots)\,,\\
	{m_{2,\psi^{(n)}}}&=\sqrt{M_{F}^{2}+\Big(\dfrac{n\pi}{L-L_{1}}\Big)^{2}}\,,&(n=1,2,\cdots)\,.
	\end{align}
%
It should be emphasized that no eigenfunctions with non-zero eigenvalues take 
non-trivial values on both $D_{1}$ and $D_{2}$.
This is because there is no degeneracy for non-zero eigenvalues i.e. 
{$m_{i,\psi^{(n)}} \neq m_{i',\psi^{(n')}}$} if $n \ne n'$ or $i \ne i'$ (except for $n=n'=0$),
as long as $L_{1}$ is not equal to $L/2$.
Hence, any linear combination of ${\mathscr{G}^{(n)}_{1,\psi_{L}}(y)}$ and ${\mathscr{G}^{(n)}_{2,\psi_{L}}(y)}$ for $n \ne 0$
\big(${\mathscr{F}^{(n')}_{1,\psi_{R}}(y)}$ and ${\mathscr{F}^{(n')}_{2,\psi_{R}}(y)}$\big) cannot become a solution to 
the eigenvalue equation (\ref{eigen-1}) ((\ref{eigen-2})).
}

The eigenfunctions ${\mathscr{G}^{(n)}_{i,\psi_{L}}(y)}$ and ${\mathscr{F}^{(n)}_{i,\psi_{R}}(y)}$ satisfy the orthonormal relations
	\begin{align}
	 \int_{D}dy \bigl({\mathscr{G}^{(n)}_{i,\psi_{L}}(y)}\bigr)^{\ast}{\mathscr{G}^{(m)}_{j,\psi_{L}}(y)}&=\delta_{n,m}\delta_{i,j},\nonumber\\
	 \int_{D}dy \bigl({\mathscr{F}^{(n')}_{i,\psi_{R}}(y)}\bigr)^{\ast}{\mathscr{F}^{(m')}_{j,\psi_{R}}(y)}&=\delta_{n',m'}\delta_{i,j},
	\label{orthonormal-relations}
	\end{align}
for {$n, m=0,1,2,\cdots$}, $n', m'=1,2,\cdots$ and $i, j=1,2$. Furthermore, they obey the {relations} (sometimes called supersymmetry relations)
	\begin{align}
	(\partial_{y}+M_{F}){\mathscr{F}^{(n)}_{i,\psi_{R}}(y)}&=m_{i,\psi^{(n)}}{\mathscr{G}^{(n)}_{i,\psi_{L}}(y)}\quad\text{on }D,\label{SUSYrelations-1}\\
	(-\partial_{y}+M_{F}){\mathscr{G}^{(n)}_{i,\psi_{L}}(y)}&=m_{i,\psi^{(n)}}{\mathscr{F}^{(n)}_{i,\psi_{R}}(y)}\,\quad\text{on }D,\label{SUSYrelations-2}
	\end{align}
for $n=0,1,2,\cdots$ and $i=1,2$ with ${\mathscr{F}^{(0)}_{i,\psi_{R}}(y)}\equiv0$.

A crucially important fact is that the set of the eigenfunctions ${\big\{}{\mathscr{G}^{(n)}_{i,\psi_{L}}(y)}; n=0,1,2,\cdots, i=1,2{\big\}}$ {\big(}${\big\{}{\mathscr{F}^{(n)}_{i,\psi_{R}}(y)}; n=1,2,\cdots, i=1,2{\big\}}${\big)} forms a complete set for square integrable functions on {$D$} with the BC (\ref{BCf}) $\bigl((\ref{BCg})\bigr)$. This is because the differential operator $-\partial_{y}^{2}+M_{F}^{2}$ is Hermitian with the BC (\ref{BCf}) $\bigl((\ref{BCg})\bigr)$ and the set of the eigenfunctions ${\big\{}{\mathscr{G}^{(n)}_{i,\psi_{L}}(y)}{\big\}}$ 
{\big(}${\big\{}{\mathscr{F}^{(n)}_{i,\psi_{R}}(y)}{\big\}}${\big)} includes all the independent eigenfunctions of (\ref{eigen-1}) $\bigl( (\ref{eigen-2}) \bigr)$ with the BC (\ref{BCf}) $\bigl((\ref{BCg})\bigr)$.

The above observation shows that the five-dimensional fields $\Psi_{L}(x,y)$ and $\Psi_{R}(x,y)$ with the BCs
	\begin{align}
	(-\partial_{y}+M_{F})\Psi_{L}(x,y)&=0\qquad \text{at}\quad y=0,L_{1}-\varepsilon, L_{1}+\varepsilon,L,\\
	\Psi_{R}(x,y)&=0\qquad \text{at}\quad y=0,L_{1}-\varepsilon, L_{1}+\varepsilon,L,
	\end{align}
can be decomposed, without any loss of generality, as
	\begin{align}
	\Psi_{L}(x,y)&=\sum^{\infty}_{n=0}\sum^{2}_{i=1}{\psi^{(n)}_{i,L}(x)\mathscr{G}^{(n)}_{i,\psi_{L}}(y)} \quad \ \text{on }D,\\
	\Psi_{R}(x,y)&=\sum^{\infty}_{{n=1}}\sum^{2}_{i=1}{\psi^{(n)}_{i,R}(x)\mathscr{F}^{(n)}_{i,\psi_{R}}(y)} \quad \text{on }D,
	\end{align}
where the coefficients of the decompositions {$\psi^{(n)}_{i,L}(x)$ and $\psi^{(n)}_{i,R}(x)$} correspond to four-dimensional left-handed and right-handed chiral fermions, respectively.

Inserting the above expansions into the five-dimensional Dirac action (\ref{eq_app:5d_action}) and integrating it over $y$ with the orthonormal relations (\ref{orthonormal-relations}) and the supersymmetry relations (\ref{SUSYrelations-1}), (\ref{SUSYrelations-2}), we find
	\begin{align}
	S = \int d^{4}x \left\{\sum^{2}_{i=1}{\overline{\psi_{i,L}^{(0)}}(x)} \bigl(i\gamma^{\mu}\partial_{\mu}\bigr){\psi^{(0)}_{i,L}(x)}+\sum^{\infty}_{n=1}\sum^{2}_{i=1}{\overline{\psi^{(n)}_{i}} (x)} \bigl(i\gamma^{\mu}\partial_{\mu}+{m_{i,\psi^{(n)}}}\bigr){\psi^{(n)}_{i}(x)}\right\}
	\label{eq_app:effective_Lagrangian}
	\end{align}
where ${\psi^{(n)}_{i}(x)\equiv \psi^{(n)}_{i,L}(x)+\psi^{(n)}_{i,R}(x)}$ for $n=1,2,\cdots$
{and} $i=1,2$.
{Here, we have used $\Gamma^{y} = -i\gamma_{5}$.} 
It follows that the four-dimensional massless left-handed chiral fermions ${ \psi^{(n)}_{i,L}}$ ($i=1,2$) appear in the four-dimensional spectrum and they are \textit{{twofold}} degenerate. The ${\psi^{(n)}_{i}}$ ($n=1,2,\cdots, i=1,2$) corresponds to a four-dimensional massive Dirac fermion with mass ${m_{i,\psi^{(n)}}}$. (For a special case of $L_{1}=L/2$, ${\psi^{(n)}_{1}}$ and ${\psi^{(n)}_{2}}$ are degenerate with the same mass ${m_{1,\psi^{(n)}}}={m_{2,\psi^{(n)}}}$, otherwise they are non-degenerate.)

Several comments are provided for completeness.
\begin{itemize}
\item
If we impose the {BCs} (instead of those in Eq.~(\ref{eq_app:BC_Psi_R_zero}))
\begin{align}
\Psi_L = 0 \quad \text{at } y = 0, \ L_1 - \varepsilon, \ L_1 + \varepsilon , {L} \,,
\end{align}
with the associated {BCs},
\begin{align}
(\partial_y + M_F) \Psi_R = 0 \quad \text{at } y = 0, \ L_1 - \varepsilon, \ L_1 + \varepsilon , {L} \,,
\end{align}
we obtain twofold-degenerated right-handed chiral zero modes (see Ref.~\cite{Fujimoto:2012wv}).
\item
A simple generalization with multiple point interactions {can be} analyzed straightforwardly.
Especially, the case with two point interactions is attractive since threefold-degenerated chiral zero modes are realized (see Ref.~\cite{Fujimoto:2012wv}).
\item
An intrinsic profile of point interaction(s) can be arranged for each 5d fermion field individually.
This property is one of the key ingredients of the flavor model proposed in Ref.~\cite{Fujimoto:2012wv}.
{At the two end points to the contrary, BCs should be arranged for all of the fields living in the bulk since the points are {kinds} of singularities on the background space.}
\item
Since we removed the singular point from the interval,
according to the formulation~\cite{Cheon:2000tq}, 
{no contribution via `brane-localized terms' emerges in integrations along the $y$ direction in the formulation.}
\item
{While no degeneracy is observed (except for the specific situation in $L_1 = L/2$) in the massive KK modes, twofold degenerated states are found as two left-handed chiral zero modes under the BCs in~(\ref{eq_app:BC_Psi_R_zero}) (and (\ref{eq_app:associated_BC})).
This observation means that the form of the zero-mode eigenfunctions ${\mathscr{G}^{(0)}_{1,\psi_{L}}}$ and ${\mathscr{G}^{(0)}_{2,\psi_{L}}}$ (shown in~(\ref{eq_app:specific_profile_first}) and (\ref{eq_app:specific_profile_first_2})) is not the unique choice.
For example, we can consider two suitable linear combinations of them, i.e., ${G^{(0)}_{1,\psi_{L}}}\equiv a \,{\mathscr{G}^{(0)}_{1,\psi_{L}}} + b \,{\mathscr{G}^{(0)}_{2,\psi_{L}}}$ and ${G^{(0)}_{2,\psi_{L}}} \equiv a' \, {\mathscr{G}^{(0)}_{1,\psi_{L}}} + b' \, {\mathscr{G}^{(0)}_{2,\psi_{L}}}$ (assuming $a, a', b, b'$ being real).
Even after imposing the orthonormality condition in~(\ref{orthonormal-relations}) in the set {$\big\{ G^{(0)}_{i,\psi_{L}}, \mathscr{G}^{(n')}_{i,\psi_{L}}; n'=1,2,\cdots, i=1,2 \big\}$}
instead of ${\big\{ \mathscr{G}^{(n)}_{i,\psi_{L}}}; n=0,1,2,\cdots, i=1,2 \big\}$, one real degree of freedom remains to be unfixed, where we obtain a series of the zero-mode eigenfunctions parametrized by the remaining real degree of freedom.
We focus on a concrete {expression,}
\begin{align}
{
\begin{pmatrix}
G^{(0)}_{1,\psi_{L}}(y) \\
G^{(0)}_{2,\psi_{L}}(y)
\end{pmatrix}
}
=
\begin{pmatrix}
\cos{\theta} & \sin{\theta} \\
-\sin{\theta} & \cos{\theta}
\end{pmatrix}
{
\begin{pmatrix}
\mathscr{G}^{(0)}_{1,\psi_{L}}(y) \\
\mathscr{G}^{(0)}_{2,\psi_{L}}(y)
\end{pmatrix}
}
\end{align}
with $0 \leq \theta \leq 2\pi$.
It is important that no difference comes out in the form of the effective Lagrangian described in~(\ref{eq_app:effective_Lagrangian}) from the original 5d action in~(\ref{eq_app:5d_action}) through KK decomposition, irrespective of the choice of the parameter $\theta$.
A clear reason behind this fact is that the free system is exactly solved (and the twofold degeneracy is found among the zero mode), where no external interaction which discriminates the difference in the zero-mode profiles is switched on.}
\\ \\
Situations {are} changed when we switch on interactions, which includes mass perturbation.
When we introduce an additional mass term as mass perturbation, and if consequently the degeneracy is resolved, no redundancy remains in the form of the mass eigenstates.\footnote{
{A good example of such mass perturbation is the introduction of the mass term from the Yukawa interaction with the $y$-dependent Higgs VEV which we discussed in {section~\ref{subsec:Mass hierarchy}}.
Before the introduction, the threefold-degenerated zero modes are massless.
However, after the introduction, the position-dependent VEV discriminates the three states and eventually the mass hierarchy is realized.}
}
Here, the rotational degree of freedom does not change the mass eigenvalues after the perturbation and does not affect physics (even though the diagonalizing matrix depends on $\theta$).
To take the simplest choice $\theta = 0$ makes analyses transparent.
\item {It is noted that we can introduce point interactions in orbifolds.
Here, we sketch how to obtain twofold degenerated localized chiral zero modes
in the geometry of $S^1/Z_2$, where we consider that the fundamental region of $y$
is $[0 + \varepsilon, L]$, which is shrunken from that of $S^1$, $[-L, L]$.
The $Z_2$ symmetry is the identification under the reflection $y \to -y$,
where 
\begin{equation}
\Psi(x, -y) =  \eta_{\Psi} \gamma_5 \Psi (x, y)
\end{equation}
is imposed with the parity $\eta_{\Psi} = \pm 1$.
Instead of (\ref{eq_app:5d_action}), we focus on the fermion action
\begin{equation}
S_{S^1/Z_2} =
\int d^4 x \left[ \int_0^{{L_1-\varepsilon}} dy 
     + \int_{{L_1+\varepsilon}}^L dy  \right]
 \overline{\Psi}(x,y) \left( i \partial_M \Gamma^M + M_F \epsilon(y) \right) \Psi(x,y)\,,
\end{equation}
where $\epsilon(y)$ represents the sign function, which is a compulsory factor to make the mass term $Z_2$ invariant (see e.g. \cite{Kaplan:2001ga}). 
Here, we select $\eta_\Psi = -1$ for realizing left-handed chiral zero mode and introduce a point interaction at $y = L_1$ which put the additional
BC for $\Psi(x,y)$ as
\begin{align}
\Psi_R = 0 \quad \text{at } y = L_1 - \varepsilon, \ L_1 + \varepsilon \,.
\end{align}
Apparently in the fundamental region $[0 + \varepsilon, L]$, the two left-handed zero modes are realized as (\ref{eq_app:specific_profile_first}) and (\ref{eq_app:specific_profile_first_2}).
If the `corresponding' point interaction is introduced at $y = - L_1$ as
\begin{align}
\Psi_R = 0 \quad \text{at } y = - L_1 + \varepsilon, \ - L_1 - \varepsilon \,,
\end{align}
the whole system remains to be $Z_2$ symmetric.}
\end{itemize}

\bibliographystyle{utphys}
\bibliography{references}

\end{document}